\documentclass[aps,reprint,prd,nofootnoteinbib]{revtex4-1}

\usepackage{bm}
\usepackage{graphicx} %
\usepackage{amsmath} %
\usepackage{amssymb} %
\usepackage{url}
\usepackage{subfigure} %
\usepackage{float} %
\usepackage{upgreek} %
\usepackage{multirow} %
\usepackage{color} %
\usepackage{lineno} %
\usepackage{hyperref} %
\usepackage{booktabs}
\usepackage{placeins}
\usepackage{ulem}
\usepackage[usenames,dvipsnames]{xcolor}
\hypersetup{colorlinks=true, urlcolor=black, linkcolor=blue, citecolor=red} %

\def\LCDM{$\Lambda$\rm{CDM}}

\usepackage{amsmath}
\usepackage{tcolorbox}
\allowdisplaybreaks[4]

\begin{document}

\title{Modeling iterative reconstruction and displacement field in the large scale structure}

\author{Atsuhisa Ota${}^1$}
\email{ota@ohio.edu}
\author{Hee-Jong Seo${}^1$}
\author{Shun Saito${}^{2,3}$}
\author{Florian Beutler${}^{4}$}
\affiliation{${}^1$Department of Physics and Astronomy Ohio University, Athens, OH, 45701, USA}
\affiliation{${}^{2}$ Institute for Multi-messenger Astrophysics and Cosmology, Department of Physics\\
Missouri University of Science and Technology, 
1315 N. Pine St., Rolla MO 65409, USA}
\affiliation{${}^{3}$Kavli Institute for the Physics and Mathematics of the Universe (WPI), Todai Institutes for Advanced Study,\\
the University of Tokyo, Kashiwanoha, Kashiwa, Chiba 277-8583, Japan}
\affiliation{${}^{4}$Institute for Astronomy, University of Edinburgh, Royal Observatory, Blackford Hill, Edinburgh EH9 3HJ, UK}

\date{\today}

\begin{abstract}
The next generation of galaxy surveys like the Dark Energy Spectroscopic Instrument (DESI) and Euclid will provide datasets orders of magnitude larger than anything available to date. Our ability to model nonlinear effects in late time matter perturbations will be a key to unlock the full potential of these datasets, and the area of initial condition reconstruction is attracting 
growing attention.
Iterative reconstruction developed in Ref.~\cite{Schmittfull:2017uhh} is a technique designed to reconstruct the displacement field from the observed galaxy distribution.
The nonlinear displacement field and initial linear density field are highly correlated. Therefore, reconstructing the nonlinear displacement field enables us to extract the primordial cosmological information better than from the late time density field at the level of the two-point statistics.
This paper will test to what extent the iterative reconstruction can recover the true displacement field and construct a perturbation theory model for the postreconstructed field. We model the iterative reconstruction process with Lagrangian perturbation theory~(LPT) up to third order for dark matter in real space and compare it with $N$-body simulations. 
We find that the simulated iterative reconstruction does not converge to the nonlinear displacement field, and the discrepancy mainly appears in the shift term, i.e., the term correlated directly with the linear density field. On the contrary, our 3LPT model predicts that the iterative reconstruction should converge to the nonlinear displacement field. We discuss the sources of discrepancy, including numerical noise/artifacts on small scales, and present an ad hoc phenomenological model that improves the agreement.

 

\end{abstract}

\maketitle


\section{Introduction}
Ongoing and upcoming large galaxy surveys~\cite{Aghamousa:2016zmz,Laureijs:2011gra,Hill:2008mv,Adams:2010un,Dawson:2015wdb,Ellis:2012rn,Spergel:2013tha} provide us a wealth of information about the initial condition of density perturbations. However, accurately modeling the low redshift matter distribution is a challenge in modern cosmology, as the history of structure formation suffers from the nonlinear evolution of density fields. Although various improvement has been made for perturbation theories~\cite{10.1093/mnras/203.2.345,Fry:1983cj,Suto:1990wf,Makino:1991rp,Jain:1993jh,Scoccimarro:1996se,Bernardeau:2001qr,Crocce:2005xy,McDonald:2006hf,McDonald:2006hf,Crocce:2007dt,Taruya:2007xy,Matarrese:2007aj,Matarrese:2007wc,Takahashi:2008yk,Bernardeau:2008fa,Carlson:2009it,Shoji:2009gg,Matsubara:2007wj,Pietroni:2011iz,Tassev:2011ac,Taruya:2012ut,Sugiyama:2013mpa}, small scale matter power spectrum is still difficult to model without additional fitting parameters~\cite{Baumann:2010tm,Carrasco:2012cv,Pajer:2013jj,Manzotti:2014loa,Carroll:2013oxa,Porto:2013qua}.
$N$-body simulations that numerically solve the evolution of all representative matter particles in the Universe can be a solution.
However, this is computationally expensive, making it challenging to browse all possible parameter spaces to find the best fit cosmological parameters to the observed data.
Alternatively, a method to reduce the nonlinear effect in the observed galaxy distribution can potentially ease modeling the late time field and extracting cosmological information from large galaxy surveys.

Baryon acoustic oscillation~(BAO) reconstruction has been developed in this context~\cite{Eisenstein:2006nk}, and a series of both numerical and analytic investigations have been conducted to understand and utilize the method~\cite{Padmanabhan:2008dd,Noh:2009bb,Tassev:2012hu,White:2015eaa,Schmittfull:2015mja,Seo:2015eyw,Hikage:2017tmm,Schmittfull:2017uhh,Chen:2019lpf,Hada:2018fde,Hada:2018ziy}.
The standard BAO reconstruction pioneered by Ref.~\cite{Eisenstein:2006nk} showed that shifting the observed particles along the gradient of the smoothed density efficiently recovers the linear BAO signals.
The standard reconstruction works because we reduce the degradation effect due to the free streaming by minimizing displacement that the particles traveled by moving them back.
The reconstructed field shows an augmented correlation with the initial field, but it is not quite the same as the initial linear density field because it cannot fully reverse all the nonlinearities~\cite{Padmanabhan:2008dd,Schmittfull:2015mja}. 
In other words, the BAO reconstruction partially brings the cosmological information of higher-order statistics back to the two-point statistics, making the reconstructed field closer to be Gaussian~\cite{Schmittfull:2015mja}.
The resulting discrepancy returns the broadband spectrum and redshift space distortion~(RSD) that deviates from the linear prediction and the prediction of the nonlinear density field. 
This procedure introduces a challenge from the modeling perspective,
consequently making it difficult to combine the BAO analysis with other large-scale structure analyses. Recently, there have been a few promising studies that address the challenge and construct perturbation theory models of postreconstruction full clustering~\cite{White:2015eaa,Hikage:2017tmm,Chen:2019lpf}. In parallel, there have been many studies on initial density field reconstruction by forward modeling~\cite{Lavaux:2019fjr,Schmidt:2020viy,Nguyen:2020yuc,Seljak:2017rmr,Feng:2018for,Modi:2021acq}. However, we should also note that, in principle, we cannot recover the initial field without solving the equation of motion~(EoM) backward, which is not reasonably possible due to the shell crossing. 
Then a natural question arises: can we find a more consistently defined late time field that is more correlated with the initial condition and less degraded than the density perturbations and try to model the field rather than reconstruct such a patchy initial density field?

\begin{figure}
  \includegraphics[width=\linewidth]{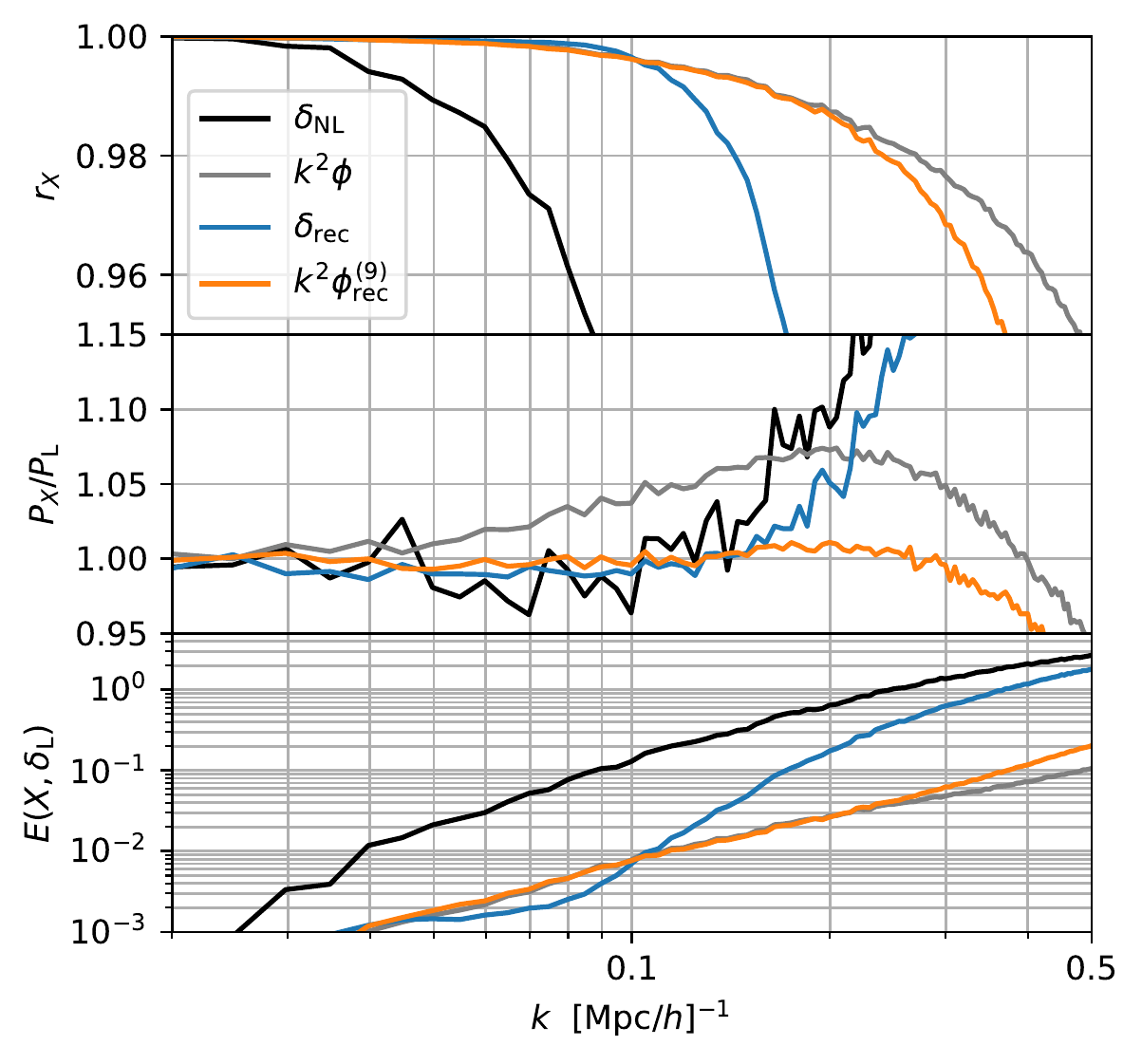}
  \caption{\textit{Top}: The cross correlation coefficient $r_X\equiv P_{X\delta_{\rm L}}/\sqrt{P_X P_{{\rm L}}}$ for the nonlinear density field $\delta_{\rm NL}$, nonlinear displacement field potential $\phi$, standard reconstruction estimator $\delta_{\rm rec}$ and 9th step iterative reconstruction estimator $\phi_{\rm rec}^{(9)}$.
   The nonlinear displacement field is 95\% correlated with the linear field for $k{\rm Mpc}/h\lesssim 0.5$, which is better than that for the standard density field reconstruction for $R=20$Mpc/$h$.
   \textit{Middle}: power spectrum of the nonlinear fields normalized by the linear matter power spectrum $P_{\rm L}$. The shift term is not reconstructed well in the iterative reconstruction.
   \textit{Bottom:} $E(X,\delta_{\rm L})$ is the power spectrum of the error field, which is the difference between the nonlinear/reconstructed displacement field and $\delta_{\rm L}$ normalized by $P_{\rm L}$~(see Sec.~\ref{simsec} and \ref{sec5} for the details of simulations.).}
  \label{fig00}
\end{figure}

In this work, we argue that the Lagrangian displacement field can be an alternative to reconstructing matter density perturbations. As our $N$-body simulations show in the top panel of Fig.~\ref{fig00}, the late time displacement divergence and initial linear density field are indeed 99\% correlated for $k{\rm Mpc}/h\lesssim 0.2$ and 95\% for $k{\rm Mpc}/h \lesssim 0.5$ at $z=0.6$~\cite{Baldauf:2015tla}~(see Sec.~\ref{simsec} and Sec.~\ref{sec5} for the details of our simulations.). 
Moreover, the top and middle panels imply that the power spectrum is amplified rather than damped. 
Thus, both the nonlinear instability and degradation effect are reduced for the displacement field compared to the density field, and we can potentially extract a wealth of information from the displacement field if they can be precisely measured and accurately modeled.
In practice, measuring the displacement fields for actual surveys is nontrivial because we do not know the initial galaxy positions a priori.
Hence, we are interested in reconstructing the displacement field from the observed galaxy distributions, which was the motivation of Ref.~\cite{Schmittfull:2017uhh}, while their primary focus was the BAO feature.  
If we could adequately reconstruct the displacement field, the broadband power spectrum would also be consistently reconstructed. In this paper, we construct a model for the broadband power spectrum for the reconstructed displacement field.

While there have been a few comparable approaches developed to iteratively reconstruct the nonlinear displacement field~\cite{Zhu:2016xyy,Zhu:2016sjc,Hada:2018fde}, the ``iterative reconstruction'' introduced by Ref.~\cite{Schmittfull:2017uhh} iteratively moves the particles along the density gradient by progressively reducing the smoothing radius step by step until we get the almost zero density field.
The final position is an estimated Lagrangian position, and the displacement from the original, i.e., the observed Eulerian position, is the reconstructed displacement field.
Ref.~\cite{Schmittfull:2017uhh} simulated the postreconstruction displacement field and showed that it is 95$\%$ correlated with the initial linear density up to $k{\rm Mpc}/h=$0.35 at $z$=0.

Although the top panel in Fig.~\ref{fig00} shows the linearity of both the displacement field and post-iterative reconstruction field, it does not necessarily mean that we can model those fields within the linear theory.
The displacement field power spectrum has about an 8\% discrepancy from the linear matter field at $k{\rm Mpc}/h=0.2$ due to the shift term, i.e., the term correlated directly with the linear density field as shown in the middle panel of Fig.~\ref{fig00}.
In addition, the post-iterative reconstruction estimator does not perfectly recover the true displacement field even for the BAO scale.
On the other hand, the bottom panel shows that the power spectrum of the errors from the linear matter field is similar to the nonlinear/reconstructed displacement field despite the discrepancy in the power spectrum in the middle panel.
Thus, the relation among those fields is not simple.

The authors of Ref.~\cite{Schmittfull:2017uhh} called the displacement field reconstruction ``$\mathcal O(1)$'' reconstruction, and they further reconstruct the initial density field from the reconstructed displacement field perturbatively, calling it ``$\mathcal O(2)$''. 
This paper takes a rigorous approach to directly model the ``$\mathcal O(1)$'' estimator using perturbation theory.
We present a theoretical modeling of the reconstruction estimator in Lagrangian perturbation theory~(LPT) up to third order for dark matter without redshift space distortions just for simplicity.
Ref.~\cite{Schmittfull:2017uhh} presented a motivation for perturbation theory modeling for their iterative reconstruction.
However, they did not consider modeling the iterative steps but provided a way to find the Zel'dovich displacement from the measured density.
This paper will model the iterative steps explicitly and find $n$-th step displacement field up to 1-loop order perturbations.

This paper is organized as follows.
Sec.~\ref{simsec} provides the details of our two simulations and their measurements.
In Sec.~\ref{sec2}, we explain the difficulty of perturbation theory for the displacement field towards an application to the iterative reconstruction.
Sec.~\ref{moderec} provides the reconstruction algorithm in the LPT perspective.
In, Sec.~\ref{sec4}, we present our LPT modeling of the iterative reconstruction and give a comparison with the simulations in Sec.~\ref{sec5}.
Then we give a summary and conclusions in the last section.

\section{Measurement of the simulated displacements}\label{simsec}

\begin{table*}
\caption{\label{table1}Simulation and sampling parameters. 
Both simulations assume a flat $\Lambda$CDM cosmology based on Ref.~\cite{Ade:2015xua} with $\Omega_{\rm m} = 0.3075$, $\Omega_{\rm b}h^2=0.0223$, $h=0.6774$, and $\sigma_8=0.8159$.}
\begin{ruledtabular}
\begin{tabular}{lccccc}
Name & subsampling \% & \# of meshes  & Box size [Mpc$/h]^3$ & \# of particles &\# of simulations \\
\hline
L500 &4 & $512^3$  & $500^3$ & $1536^3$ & 5\\
subL500 &0.15 & $512^3$  & $500^3$ & $1536^3$ & 5\\
L1500 &4 	 & $1024^3$ & $1500^3$ & $1536^3$ &1  \\
fullL1500 &100  & $1536^3$  & $1500^3$ & $1536^3$ &1 \\
\end{tabular}
\end{ruledtabular}
\end{table*}

To compare our theoretical model with realistic nonlinear/reconstructed displacement fields, we need to minimize nonphysical numerical artifacts in measuring the simulated displacement fields. 
In this section, we explain the setup of our numerical simulations.
Readers who are only interested in the theoretical implementation can skip this section.

\medskip
Measuring the displacement field or velocity field is challenging due to these vector fields' discrete and nonuniform sampling. 
Measuring the displacement using mass tracers gives the mass-weighted displacement field, giving no measurement in the locations without mass, while perturbation theories derive a volume-weighted quantity. 
The Delaunay tessellation method (e.g., Ref.~\cite{Pueblas:2008uv}) works for a volume-weighted measurement, which is more robust to the discreteness effect. 
In this paper, however, we adopt the mass-weighted measurement, following the convention of Ref.~\cite{Schmittfull:2017uhh}, for both calculating the Lagrangian displacement field and the reconstructed displacement field.
As a caveat, one could instead directly model the mass-weighted quantity in perturbation theory~\cite{Vlah:2012ni}.
We test the sanity of the mass-weighted measurement using a convergence test as a function of various levels of discreteness effect against a reference simulation without the discreteness effect. 
The mass-weighted displacement can be written as
\def \bx{\mathbf x}
\begin{align}
	\mathbf \Psi^{\rm obs.}_p 
	&= \frac{\sum_i W_{\rm CIC}({\bx_p,\bx_i}) \mathbf \Psi^{\rm obs.}(\bx_i)  }{\sum_i W_{\rm CIC}({\bx_p,\bx_i}) }
	,\label{mweight}
\end{align}
where $\mathbf \Psi^{\rm obs.}(\bx_i)$ is the observed displacement field of $i$-th particle, $W_{\rm CIC}$ is the pixel window function indicating that we are using the Cloud-in-Cell assignment, and $\mathbf \Psi^{\rm obs.}_p$ is the mass-weighted displacement field for a pixel centered at $\bx_p$. For pixels with no particles found, we incorrectly set $\mathbf \Psi^{\rm obs.}_p=0$. 
This operation biases the result; the overall amplitude is reduced by the fraction of the zeroed pixels, which we correct with a simple multiplicative rescaling. However, this prescription also introduces a spurious small-scale power correlated with the pixel window function and the nonuniform particle spacing, which we cannot straightforwardly correct. Note that $\mathbf \Psi$ is evaluated at the initial Lagrangian position, which is fixed at $\mathbf x_{p}$, and that this is fundamentally different from estimating the velocity divergence field, which we evaluate in the Eulerian position.

We show the subsampling parameters~(i.e., particle spacing) and pixel window function of our two numerical simulations in Tab.~\ref{table1}. Simulations assume a flat \LCDM~cosmology based on Ref.~\cite{Ade:2015xua} with $\Omega_{\rm m} = 0.3075$, $\Omega_{\rm b}h^2=0.0223$, $h=0.6774$, and $\sigma_8=0.8159$. 
This is the same cosmology used in Ref.~\cite{Schmittfull:2017uhh}.  
Full $N$-body simulations were produced using the MP-Gadget~\cite{Springel:2000yr,Springel:2005mi,2018zndo...1451799F} with the box size of $500{\rm Mpc}/h$ and $1500{\rm Mpc}/h$. For $500{\rm Mpc}/h$, we use the average of five simulations. Both simulations use a particle resolution of $1536^3$. The simulation evolves $1536^3$ particles from $z=99$ by computing forces in a grid of $1536^3$, and we subsample 4\% of the output particles at $z=0.6$ to make L1500 (for a box of $1500{\rm Mpc}/h$) and L500 (for a box of $500{\rm Mpc}/h$). We use a grid of $512^3$ to Fourier-transform, reconstruct this nonlinear field for L500 and subL500, and use a grid of $1024^3$ for fullL1500 and L1500.

Here, fullL1500 is the simulation of $1500{\rm Mpc}/h$ without subsampling, therefore uniformly being distributed in the Lagrangian position. The mass-weighted displacement for this simulation should be equivalent to the volume-weighted measurement, and we expect no discreteness and no pixel window function effect for this set. We do not use this complete catalog for reconstruction, as, first, it is computationally expensive for iterative steps, and second, the reconstructed density field will suffer the discreteness and the pixel window function effect even if we used this complete set, as we could not perfectly recover the Lagrangian, uniform position even after reconstruction.
We use this complete set as our reference to estimate the convergence of the simulation L500, which is our main set to test the broadband shape after reconstruction.
We find that the displacement field of the two simulations is convergent up to  $k{\rm Mpc}/h\lesssim 0.2$ for those mesh resolutions within 1.1\%. We find that further subsampling and/or increasing the mesh size breaks this convergence.
Given the mesh resolution and the particle resolution of these two simulations in  Tab.~\ref{table1},  we consider 1 ${\rm Mpc}/h$ as the limiting resolution and account for it in our LPT model.
The convergence of the displacement field does not straightforwardly imply the convergence of the post-iterative reconstruction displacement field.  
However, in this paper, we shelve the plans for 100\% sampling for the iterative reconstruction mainly because of our numerical resources.
To see the numerical stability of the iterative reconstruction, we also introduced subL500, which is a 0.15\% sampling of the $512^3$ simulation and the rest of the parameters are the same as those for L500.

For the BAO feature comparison, L500 is not optimal due to its small box size, i.e., $500{\rm Mpc}/h$.
For this reason, we apply the iterative reconstruction on a pair of L1500, the one generated with an initial condition with BAO and the one generated without BAO with the same white-noise fields~\cite{Wojtak:2016sxr,Schmittfull:2017uhh,Ding:2017gad}, which will be shown in Fig.~\ref{fig2}. By pairing and dividing, we cancel out the cosmic variance and the spurious effect due to the discreteness aforementioned.

As a caveat, note that the observed field from galaxy surveys will be subject to many orders of magnitude more severe discreteness effects than our default simulation L500. Our companion paper~\cite{Seo:2021} discusses a surrogate reconstruction method to avoid the discreteness in reconstruction. 
In this paper, however, we focus on modeling the reconstructed displacement in the shot-noiseless limit without redshift space distortion to test the model with minimal complications.

\section{The UV mistake of 3LPT displacement}\label{sec2}

This paper aims at modeling the iterative reconstruction up to 1-loop order, which can also be referred to as ``3LPT'' as we expand the equation of motion up to third order in the linear density field.
We also call the Zel'dovich approximation, i.e., linear perturbations, ``1LPT'' in this paper.

One might wonder if the perturbation theory for the displacement field is more convergent and more manageable than that for the density field, since we mentioned that the simulated nonlinear displacement field is more correlated with the linear field than the density field.
However, it does not work as we expected because of the ``UV-mistake'', which we explain in this section.
The UV mistake, first pointed out by Ref.~\cite{Baldauf:2015tla}, is an issue about the cut-off dependence of a loop calculation for the displacement field.

\medskip
Let $\mathbf x$ and $\mathbf q$ be the Eulerian and Lagrangian coordinates.
The relation between these two coordinates is written as 
\begin{align}
	\mathbf x = \mathbf q + \mathbf \Psi(\mathbf q),\label{psidef}
\end{align}
and $\mathbf \Psi$ is called the displacement field.
The Eulerian density perturbation is exponentiated by the Lagrangian displacement field as~\cite{Matsubara:2007wj}
	\begin{align}
   \delta_{\rm NL}(\mathbf x)
    &=  \int \frac{d^3kd^3 q}{(2\pi)^3}  e^{i\mathbf k\cdot (\mathbf x-\mathbf q)} \left [ e^{ -i\mathbf k\cdot \mathbf \Psi(\mathbf q)} 
    -
    1\right].\label{14}
\end{align}

\begin{figure}
  \includegraphics[width=\linewidth]{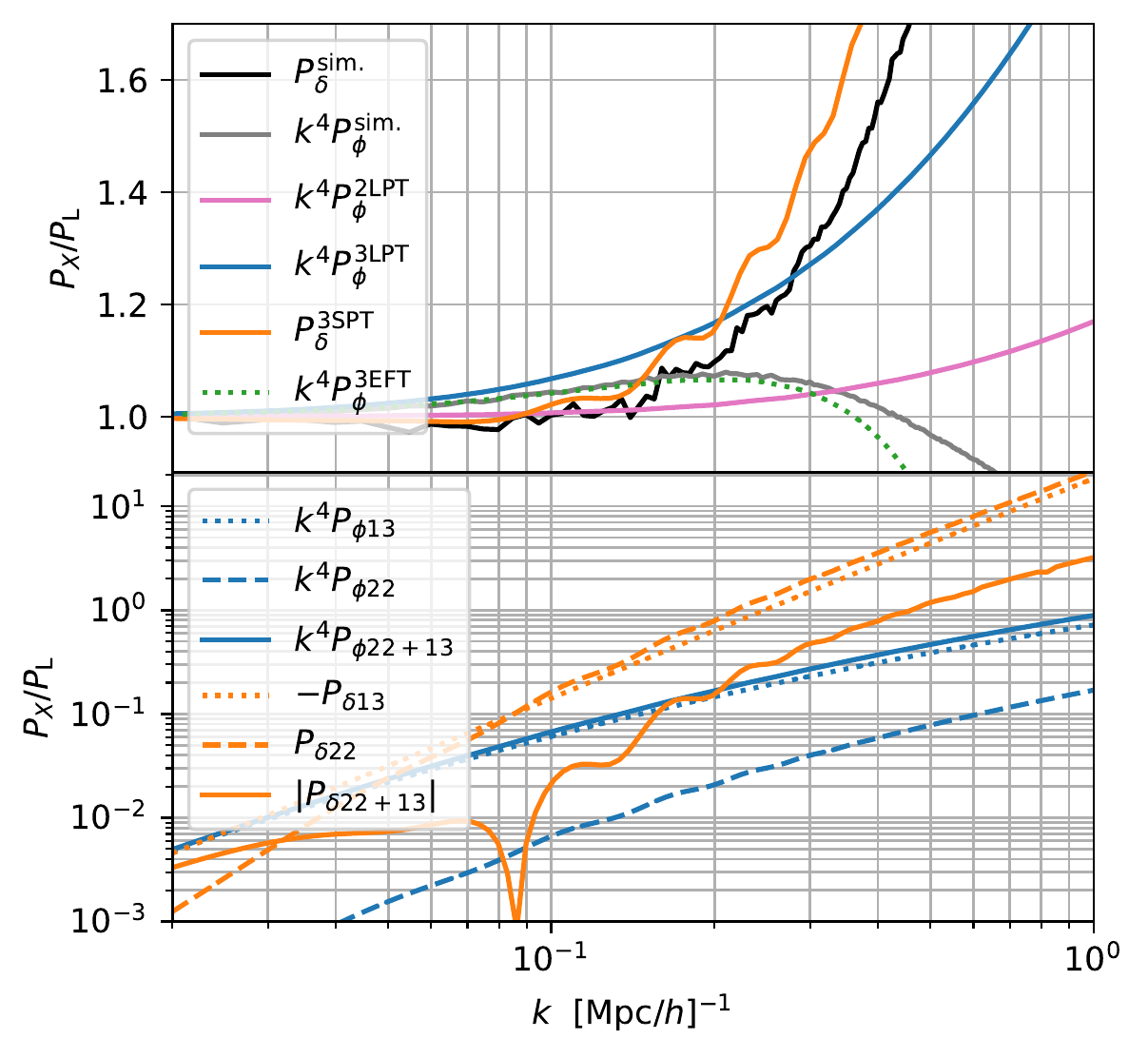}
  \caption{\textit{Top}: Comparison of the simulation displacement~(gray) and density field~(black), the 1-loop displacement power sepctrum~(blue) and 1-loop density power spectrum~(orange). We evaluated at $z$=0.6. The second order LPT is shown in pink.
  The EFT~(green) is defined in Eq.~\eqref{EFT1}.
  All power spectra are normalized by the linear matter power spectrum to better illustrate the difference.
    \textit{Bottom}: 1-loop corrections for the density~(SPT) and displacement fields~(LPT) are shown. While $|P_{\delta13}| \sim P_{\delta22} \gg k^4P_{\phi13}\gg k^4P_{\phi22}$, we have $P_{\rm L}\sim P^{\rm 3SPT}_{\delta}<k^4P^{\rm 3LPT}_{\phi}$ for $k{\rm Mpc}/h<0.1$, and thus the low $k$ displacement field is sensitive to the high $k$ loop contributions.}
  \label{fig0}
\end{figure}

\medskip
We introduce the displacement field potential $\phi$ that satisfies $i\mathbf k\phi =  \mathbf \Psi$ in Fourier space.
The 1LPT spectrum of $\phi$ is straightforwardly written by the linear matter power spectrum $P_{\rm L}$:
\begin{align}
    P^{\rm 1LPT}_{\phi}=& \frac{P_{\rm L}}{k^4},
\end{align}
Then, the 3LPT power spectrum is given as
\begin{align}
    P^{\rm 3LPT}_{\phi}=& P^{\rm 1LPT}_{\phi}+ P_{\phi13}+ P_{\phi22} \label{LPT:disp},
\end{align}
where we defined~\cite{Baldauf:2015tla}
\begin{align}
P_{\phi22}=& \frac{9}{98}\int  dxd\mu \frac{P_{\rm L}(kx)P_{\rm L}(ky)}{4\pi^2k}
		 \frac{x^2(1-\mu^2)^2}{y^4},\label{P22}\\
P_{\phi13}=&
\frac{10}{21}P_{\rm L}(k)\int  dxd\mu  \frac{P_{\rm L}(k x)}{4\pi^2k}   
		 \frac{x^2(1-\mu^2)^2}{y^2 },
		\label{P13}
\end{align}
with $y\equiv(1 - 2x \mu+x^2)^{1/2}$.
The integral variables $x$ and $\mu$ run from $0$ to $\infty$ and $-1$ to 1 respectively.
We show $P^{\rm 3LPT}_{\phi}$, $P_{\phi13}$ and $P_{\phi22}$ and their standard perturbation theory~(SPT) counterpart for the density field in Fig.~\ref{fig0}~(see e.g., Ref.~\cite{Bernardeau:2001qr} for the SPT 1-loop term).
The 3LPT correction is dominated by $P_{\phi13}$, which is about 10\% correction to the linear power spectrum even at the BAO scale.
We explain why $P_{\phi13}\gg P_{\phi22}$ as follows.
In the large loop momentum $x\to \infty$ limit, we get~\cite{Baldauf:2015tla}
\begin{align}
	k^4P_{\phi22}\to & \frac{24}{245}\frac{k^3}{4\pi^2}\int  \frac{dx}{x^2} P_{\rm L}(kx)^2\, ,\\
k^4P_{\phi13} \to &
		 		 \frac{32}{63}P_{\rm L}(k)\frac{k^3}{4\pi^2}\int dx P_{\rm L}(k x) \label{UVmistake}\, ,
\end{align}
and the integral in Eq.~\eqref{UVmistake} is sensitive to the large $x$ contributions.
As we see the gray curve in Fig.~\ref{fig0}, $P_{\phi}^{\rm 1LPT}$ overestimates the displacement field at high $k$, and thus the loop integral contains the incorrect UV modes.
As a result, $P_{\phi13}$, i.e., the shift term is overestimated.
This is why the authors in Ref.~\cite{Baldauf:2015tla} called this problem ``UV-mistake''.
The UV mistake around the BAO scale is specific to the displacement field.
While we have the integral like Eq.~\eqref{UVmistake} in the 3SPT corrections, they are canceled, and the net correction is suppressed.
On the other hand, the fact that $|P_{\phi 13}|\gg |P_{\delta13}|$, while $P_{\phi 13}$ is positive and $P_{\delta13}$ is negative, and $P_{\phi 22}\ll P_{\delta22}$ is the reason that the displacement field is correlated with the linear field very well and that the phase shift in the BAO is negligible for the displacement field.

\medskip
Why do we have the UV mistake?
In the LPT, we solve the following equation of motion (EoM)~\cite{Bernardeau:2001qr}:
\begin{align}
	\frac{d^2 \mathbf \Psi}{d\eta^2} - \mathcal H \frac{d\mathbf \Psi}{d\eta} &= -\nabla_{ x} \phi_{\rm N}(\mathbf q+\mathbf \Psi),\label{eom1}
\end{align}
where $\mathcal H$, $\phi_{\rm N}$, and $\eta$ are the conformal Hubble parameter, the gravitational potential, and the conformal time, respectively.
This equation is based on Newton's equation of a point particle in the gravitational potential.
To close the EoM, we impose the Poisson equation
\begin{align}
	\nabla_x^2 \phi_{\rm N}  &= \frac{3}{2}\mathcal H^2 \delta_{\rm NL}.\label{HC}
\end{align}
The RHS of Eq.~\eqref{HC} assumes the pressure-less perfect fluid approximation, which ignores the gravitational interactions between the individual fluid elements.
The UV mistake is a consequence of this simplification, i.e., gravitational scattering of the individual fluid elements happens at a small scale, which affects the large-scale configuration of the elements. Those effects are canceled out to some extent for the density field but are not for the displacement.
$N$-body simulations (or more advanced hydrodynamical simulations) can directly account for these effects, but their analytic implementation is generally complicated and impossible.
One solution to this problem is to use an effective field theory, where we introduce all possible higher-order gradient corrections to the energy-momentum tensor and modify Eq.~\eqref{HC} to include the many-body effects.
The 1-loop level effective displacement field is written as~\cite{Porto:2013qua}
\begin{align}
	\mathbf \Psi_{\rm eff.} = (1+\alpha k^2  )\mathbf  \Psi_{(1)} + \mathbf  \Psi_{(2)} + \mathbf \Psi_{(3)} +\cdots,\label{tobias_eft}
\end{align}
and $\alpha \sim -1$ at $z=0.6$ is read from Fig.~11 in Ref.~\cite{Baldauf:2015tla}, where a subscript means the order of the perturbations.
Note that we take the upper bound of the loop integral $k_{\rm max}$ as $k_{\rm max}/k = \infty$.
In this paper, we fit $\langle \delta_{\rm L}\phi_{\rm sim.} \rangle/\langle \delta_{\rm L}^2 \rangle $ with
\begin{align}
	f(k;\alpha)
	 \equiv 1+ \alpha k^2 +  \frac{P_{\phi13}}{2P^{\rm 1LPT}_\phi} ,\label{EFT1}
\end{align}
using the data in $0.02\lesssim k{\rm Mpc}/h\lesssim 0.2$, and get $\alpha\sim-1.27$, which is slightly different from $\alpha$ in Ref.~\cite{Baldauf:2015tla}.
The sample variance can be canceled by dividing the nonlinear spectrum by the initial realizations in Eq.~\eqref{EFT1}.
The advantage of fitting the cross power spectrum instead of the auto power spectrum is to avoid overfitting noise and $P_{\phi22}$.
Our EFT is shown in Fig.~\ref{fig0} as the green dotted curve.
Note that the authors in the reference got $\alpha$ by fitting the data at the field level while we got them at the power spectrum level for simplicity.
The comparison of these two approaches and the overfitting issue are discussed in their paper.

\medskip
To summarize, the 3LPT, mainly the shift term, suffers the ``UV mistake,'' which requires correcting the mistake beyond the perturbation theory framework.
Do we also have the same problem with the LPT reconstruction? 
Indeed, we will show that the iterative procedure partly cancels the UV mistake, and the postreconstruction displacement field agrees with the true displacement field up to much higher $k$.

\section{Displacement field reconstruction}\label{moderec}

In this section, we review the standard and iterative reconstruction algorithm from the viewpoint of the Lagrangian perturbation theory, following Ref.~\cite{Schmittfull:2017uhh}

\subsection{Standard reconstruction algorithm}\label{stdbao}

The standard density field reconstruction algorithm is summarized as follows~\cite{Eisenstein:2006nk,Padmanabhan:2008dd}: 
\begin{enumerate}
	\item Firstly, measure the density field $\delta_{\rm NL}$ in the Eulerian frame $\mathbf x$. The Lagrangian expression of the nonlinear density field is given in Eq.~\eqref{14}.
The Fourier transform of the density field is then
	  \begin{align}
     \delta_{\rm NL}(\mathbf k)
    &=   \int d^3 qe^{-i\mathbf k\cdot \mathbf q} \left [ e^{ -i\mathbf k\cdot \mathbf \Psi(\mathbf q)} 
    -
    1\right].\label{17}
\end{align}

	\item Smooth the density field as $  \delta_{\rm NL}(\mathbf k) \to S(k)  \delta_{\rm NL}(\mathbf k)$. $S(k)$ is a smoothing window in Fourier space, and we usually choose a Gaussian window. We smooth the density field to remove the nonlinear part of the density field to use the linear continuity equation to estimate the displacement.  
	\item Introduce the shift vector as the gradient of the smoothed density field. In Fourier space this operation is equivalent to 
	\begin{align}
		  {\mathbf s}(\mathbf k) \equiv  \frac{-i\mathbf k}{k^{2}}S(k)  \delta_{\rm NL}(\mathbf k).\label{shiftdef}
	\end{align} 
	\item Introduce the displaced density field 
	\begin{align}
	   \delta_{\rm d}(\mathbf k) \equiv    \int d^3 qe^{-i\mathbf k\cdot \mathbf q} \left [ e^{ -i\mathbf k\cdot (\mathbf \Psi(\mathbf q)+\mathbf s(\mathbf x))} 
    -
    1
    \right].\label{23}
\end{align}
	$\mathbf s(\mathbf x)$ is the smoothed linear negative displacement and the linear part of the original displacement field is partly subtracted.
	Note that the above Fourier transform is with respect to $\mathbf q$, while the shift vector in the exponential is defined at $\mathbf x$~\cite{Schmittfull:2015mja}.
\end{enumerate}
In the standard reconstruction, we also introduce the shifted uniform/reference field 
\begin{align}
	   \delta_{\rm s}(\mathbf k) \equiv    \int d^3 qe^{-i\mathbf k\cdot \mathbf q} \left [ e^{ -i\mathbf k\cdot \mathbf s(\mathbf q)} 
    -
    1
    \right],
\end{align}
and the reconstruction estimator is defined as 
\begin{align}
	\delta_{\rm rec} \equiv \delta_{\rm d} - \delta_{\rm s}.
\end{align}
Expanding the estimator with respect to $\mathbf \Psi$, we get~\cite{Padmanabhan:2008dd}
\begin{align}
	\delta_{\rm rec} = -i\mathbf k\cdot \mathbf \Psi(\mathbf k) + \mathcal O(\Psi^2).
\end{align}
Ref.~\cite{Padmanabhan:2008dd} showed that we could not remove the pure second-order contribution in $\mathcal O(\Psi^2)$. Therefore, $\delta_{\rm rec}$ does not coincide with the linear field at nonlinear order.

\subsection{Iterative reconstruction algorithm}\label{itrbao}

In the above first step reconstruction, we reduce the original displacement $\mathbf \Psi(\mathbf q)$ to $\mathbf \Psi(\mathbf q)+\mathbf s(\mathbf x)$.
In Ref.~\cite{Schmittfull:2017uhh}, they repeat the above steps 1 to 4 by updating the nonlinear field by the displaced density field until we get almost zero displacement in the end.
Hereafter, we put indices in the above equations as
\begin{align}
    \mathbf \Psi \to \mathbf \Psi^{(0)},~S\to S^{(0)},~\mathbf s \to \mathbf s^{(0)}, ~\delta_{\rm NL}\to \delta^{(0)}_{\rm NL},
\end{align}
and generalize 0 to $n$.
We also write $ \delta^{(n)}_{\rm NL}=  \delta^{(n-1)}_{\rm d}$.
We may rephrase step 4 with respect to the displacement field as follows:
\begin{enumerate}
\setcounter{enumi}{4}
	\item[4'.] We define the new displacement field
	\begin{align}
		\mathbf \Psi^{(1)}(\mathbf q) \equiv  \mathbf \Psi^{(0)}(\mathbf q)+\mathbf s^{(0)}(\mathbf q+\mathbf \Psi^{(0)}(\mathbf q)) \label{dispupd}
	\end{align}  and update the nonlinear field as 
	\begin{align}
   \delta^{(1)}_{\rm NL}(\mathbf x)
    &\equiv    \int \frac{d^3kd^3 q}{(2\pi)^3}  e^{i\mathbf k\cdot (\mathbf x-\mathbf q)} \left [ e^{ -i\mathbf k\cdot \mathbf \Psi^{(1)}(\mathbf q)} 
    -
    1\right].
\end{align}
\end{enumerate}

We then repeat 1 to 4' until we get $\delta^{(n)}_{\rm NL}(\mathbf x) \approx 0$. The final positions are the estimated Lagrangian positions where the Eulerian density field is zero. Generalizing Eq.~\eqref{dispupd}, the $n$-th step displacement field is recursively defined as 
	\begin{align}
		\mathbf \Psi^{(n)}(\mathbf q) \equiv  \mathbf \Psi^{(n-1)}(\mathbf q)+\mathbf s^{(n-1)}(\mathbf q+\mathbf \Psi^{(n-1)}(\mathbf q)).\label{psin}
	\end{align}
	We reduce the smoothing radius $R$ step by step, and $n$-th step smoothing function is defined as
\begin{align}
	S^{(n)} \equiv & \exp\left(-\frac{R_n^2k^2}{4}\right),\\
R_n\equiv &\epsilon^{n} R.\label{Rdef}
\end{align}
	Note that the field level smoothing scale is given as $R/\sqrt{2}$ in the present convention. 
In this paper, we use $\epsilon = 1/\sqrt{2}$ unless otherwise stated.
The remaining process is summarized as follows.

\begin{enumerate}
\setcounter{enumi}{4}

\item Evaluate the total negative displacement:
\begin{align}
	\mathbf s^{(n)}_{\rm tot.}(\mathbf x) \equiv &\mathbf s^{(0)}(\mathbf q+\mathbf \Psi^{(0)}(\mathbf q)) + \mathbf s^{(1)}(\mathbf q+\mathbf \Psi^{(1)}(\mathbf q))\notag \\
	&+\cdots + \mathbf s^{(n-1)}(\mathbf q+\mathbf \Psi^{(n-1)}(\mathbf q)).\label{totn}
\end{align}
Note that $\mathbf s^{(n)}_{\rm tot.}$ is evaluated at the initial (i.e., the observed) Eulerian position $\mathbf x$, so, if the reconstruction is nearly perfect, we reconstruct the negative displacement that satisfies
\begin{align}
	\mathbf x +\mathbf s^{(n)}_{\rm tot.}(\mathbf x) = \mathbf q.
\end{align}
	\item Evaluate the negative displacement as a function of the estimated Lagrangian position, which is given by
	\begin{align}
		{\mathbf \Psi}^{(n)}_{\rm rec}(\mathbf q) = -\mathbf s^{(n)}_{\rm tot.}(\mathbf q+\mathbf \Psi(\mathbf q)).
	\end{align}  
\end{enumerate}
This ${\mathbf \Psi}^{(n)}_{\rm rec}$ is the output of the iterative reconstruction.
We usually compute the divergence of ${\mathbf \Psi}^{(n)}_{\rm rec}$ to compare it with the density fields.
This paper also refers to the displacement field's divergence as ``displacement field'' or $k^2\phi$ unless otherwise stated.

\section{Modeling the iterative reconstruction with LPT}\label{sec4}

Our modeling strategy is as follows. Firstly, we put forward a perturbative expansion ansatz of the $n$-th step displacement field in the most general way. Then, following steps presented in Sec.~\ref{stdbao} and \ref{itrbao}, we compute the displacement field at $n+1$th step.
Comparing it with the $n$th step ansatz, we derive the recurrence relation for the expansion kernels.
Readers who are only interested in the main result may skip the detailed derivation and refer to Eqs.~\eqref{1-loop:power} and \eqref{1-loop:prop}.

\subsection{Recurrence relation}
\subsubsection{Displacement field}
Let us define the potential field of the observed particles after the $n$ iterations as 
\begin{align}
	\mathbf \Psi^{(n)}(\mathbf k) & =i\mathbf k  \phi^{(n)}(\mathbf k).  \end{align}
 The curl component is at least at third order, which does not contribute to the divergence in the end.
Then the iterative reconstruction displacement estimator ${\mathbf \Psi}_{\rm rec}^{(n)}$ is defined as the difference between and the true nonlinear displacement field $\mathbf \Psi^{(0)}(\mathbf q)$ and the  $n$-th step displacement field 
\begin{align}
	{\mathbf \Psi}_{\rm rec}^{(n)}(\mathbf q) = - \mathbf s^{(n)}_{\rm tot} (\mathbf x)=  \mathbf \Psi^{(0)}(\mathbf q) - \mathbf \Psi^{(n)}(\mathbf q),
\end{align}
where we used Eq.~\eqref{psin} and Eq.~\eqref{totn}.
The reconstructed displacement potential in Fourier space is then defined as
\begin{align}
	\phi^{(n)}_{\rm rec} \equiv \phi^{(0)} - \phi^{(n)}.\label{def:rec}
\end{align}
We put forward the $n$-th step ansatz as a series expansion with respect to the zeroth step potential field $\phi = \phi^{(0)}$ as
\begin{align} 
	&\phi^{(n)}(\mathbf k) 
	=A^{(n)}(\mathbf k) \phi(\mathbf k) \notag \\
	+&
	\frac{1}{2!}\int\frac{d^3k_1d^3k_2}{(2\pi)^6}(2\pi)^3 \delta^{(3)}_{\rm D}(\mathbf k-\mathbf k_1-\mathbf k_2) 
	\notag \\
	&\times A^{(n)}(\mathbf k_2,\mathbf k_3)\phi(\mathbf k_2)\phi(\mathbf k_3) \notag \\
	 +&
	\frac{1}{3!}\int\frac{d^3k_1d^3k_2d^3k_3}{(2\pi)^9}(2\pi)^3 \delta^{(3)}_{\rm D}(\mathbf k-\mathbf k_1-\mathbf k_2-\mathbf k_3) 
	\notag \\
	&\times A^{(n)}(\mathbf k_1,\mathbf k_2,\mathbf k_3)\phi(\mathbf k_1)\phi(\mathbf k_2)\phi(\mathbf k_3),\label{eq27}
\end{align}
where, by definition, the zeroth step kernels are given as
\begin{align}
    A^{(0)}(\mathbf k) &= 1,\label{A1_0}	\\
	A^{(0)}(\mathbf k_2,\mathbf k_3) &=A^{(0)}(\mathbf k_1,\mathbf k_2,\mathbf k_3)= 0\label{A23_0}.
\end{align}
Note that Eq.~\eqref{eq27} is the most general form of the expansion at the 1-loop order.
Then, we will get $\phi^{(n+1)}$ to find the recurrence relations for $A$.
To simplify our notation, we introduce the following notation:
\begin{align} 
	\phi^{(n)}_a =A^{(n)}_a{}^b \phi_b 
	+
	\frac{1}{2}A^{(n)}_a{}^{bc}\phi_b\phi_c  +
	\frac{1}{3!}A^{(n)}_a{}^{bcd}\phi_b\phi_c\phi_d,\label{abs28}
\end{align}
where we defined
\begin{align}
    A^{(n)}_a{}^b \equiv & A^{(n)}(\mathbf k_b) (2\pi)^3 \delta^{(3)}_{\rm D}(\mathbf k_a-\mathbf k_b) ,	\\
	A^{(n)}_a{}^{bc} \equiv & A^{(n)}(\mathbf k_b,\mathbf k_c)(2\pi)^3 \delta^{(3)}_{\rm D}(\mathbf k_a-\mathbf k_b-\mathbf k_c) . \\
	A^{(n)}_a{}^{bcd} \equiv & A^{(n)}(\mathbf k_b,\mathbf k_c,\mathbf k_d)\notag \\
	&\times (2\pi)^3 \delta^{(3)}_{\rm D}(\mathbf k_a-\mathbf k_b-\mathbf k_c-\mathbf k_d),
\end{align}
and we integrate the momentum for a pair of repeated upper and lower indices.

\subsubsection{Nonlinear density}

Given a displacement field~\eqref{abs28}, we calculate the nonlinear density field from Eq.~\eqref{14}.
Eq.~\eqref{14} can be expanded into
\begin{align}
	\delta^{(n)}_{{\rm NL},a} &= X_a{}^b \phi^{(n)}_b 
	+
	\frac{1}{2}X_a{}^{bc}\phi^{(n)}_b\phi^{(n)}_c  \notag \\
	&+
	\frac{1}{3!}X_a{}^{bcd}\phi^{(n)}_b\phi^{(n)}_c\phi^{(n)}_d,\label{abs34}
	\end{align}
where we defined
\begin{align}
	X_a{}^b &=(\mathbf k_a\cdot \mathbf k_b)(2\pi)^3\delta^{(3)}_{\rm D}(\mathbf k_a-\mathbf k_b),\\
	X_a{}^{bc} &=(\mathbf k_a \cdot \mathbf k_b)(\mathbf k_a \cdot \mathbf k_c)
	(2\pi)^3\delta^{(3)}_{\rm D}(\mathbf k_a-\mathbf k_b-\mathbf k_c),\\
	X_a{}^{bcd} &= (\mathbf k_a \cdot \mathbf k_b)(\mathbf k_a \cdot \mathbf k_c)(\mathbf k_a \cdot \mathbf k_d)
	\notag \\
	&\times 
	(2\pi)^3\delta^{(3)}_{\rm D}(\mathbf k_a-\mathbf k_b-\mathbf k_c-\mathbf k_d).
\end{align}
Note that we get the 3SPT density power spectrum when we evaluate the 1-loop density power spectrum for the expansion~\eqref{abs34}, and there is no resummation for this truncation.
Substituting Eqs.~\eqref{abs28} into \eqref{abs34}, one finds
\begin{align}
	&\delta^{(n)}_{{\rm NL},a} 
	 = X_a{}^b A^{(n)}_b{}^p \phi_p 
	\notag \\
	&+
	\frac{1}{2}\left( X_a{}^bA^{(n)}_b{}^{pq}
	+
	X_a{}^{be}A^{(n)}_b{}^p  
	A^{(n)}_e{}^q 
	\right)\phi_p\phi_q  
	\notag \\
	&
	+
	\frac{1}{3!}\left( X_a{}^bA^{(n)}_b{}^{pqr}
	+
	3X_a{}^{be}A^{(n)}_b{}^p  A^{(n)}_e{}^{qr} \right.\notag \\
	& \left.+
	X_a{}^{bdf}A^{(n)}_b{}^p 
	A^{(n)}_d{}^q
	A^{(n)}_f{}^r\right)\phi_p \phi_q \phi_r. \label{deltaNLexp}
	\end{align}
	We obtained the connection between the expansion of the potential field and the nonlinear density field. 
    Our next step is to introduce the smoothing factor 
     and then the shift vector.

	\subsubsection{Shift vector}
Let us 
introduce the shift vector and find its perturbative expansion with respect to $\phi$.
We generalize Eq.~\eqref{shiftdef} for $\delta^{(n)}_{\rm NL}$ obtained in Eq.~\eqref{deltaNLexp} and find
\begin{align}
    \mathbf s^{(n)}(\mathbf k) = -\frac{i\mathbf k}{k^2}S^{(n)}(k)\delta^{(n)}_{\rm NL}(\mathbf k).
\end{align}
As a caveat, we cannot simply update the Fourier space displacement field as $\mathbf \Psi^{(n)}(\mathbf k) + \mathbf s^{(n)}(\mathbf k)$ because we are mixing the Lagrangian field $\mathbf \Psi^{(n)}(\mathbf q)$ and the Eulerian field $\mathbf s^{(n)}(\mathbf x)$ in Eq.~\eqref{psin}. 
Indeed, the Fourier transform of  Eq.~\eqref{psin} with respect to $\mathbf q$ is
\begin{align}
    \mathbf \Psi^{(n+1)}(\mathbf k) =\mathbf \Psi^{(n)}(\mathbf k) +\mathbf t^{(n)}(\mathbf k)\label{psinexetk},
\end{align}
where we introduced 
\begin{align}
    \mathbf t^{(n)}(\mathbf k) \equiv \int d^3q e^{-i\mathbf k\cdot \mathbf q}\mathbf s^{(n)}(\mathbf q+\mathbf \Psi^{(n)}(\mathbf q)).\label{deftk}
\end{align}
Then,  we expand the above equation in terms of $\mathbf \Psi$, and identify $\int d^3q e^{-i\mathbf k\cdot \mathbf q}\mathbf s^{(n)}(\mathbf q)$ with $\mathbf s^{(n)}(\mathbf k)$. Note that $\mathbf q$ in this integral is nothing but a dummy variable. 
Then, Eq.~\eqref{deftk} is expanded into
\begin{align}
    \mathbf Y_a{}^s\delta^{(n)}_{{\rm NL}s}
    +
    \mathbf Y_a{}^{bs} \phi^{(n)}_b  \delta^{(n)}_{{\rm NL}s}
    +
    \frac{1}{2}\mathbf Y_a{}^{bcs}  \phi^{(n)}_b \phi^{(n)}_c \delta^{(n)}_{{\rm NL}s},
\end{align}
where the expansion kernels are given as
\begin{align}
	\mathbf Y_a{}^{b}&=
     \frac{-i\mathbf k_b}{k_b^2} S^{(n)}(k_b)(2\pi)^3\delta^{(3)}_{\rm D}(\mathbf k_a-\mathbf k_b),\\
	\mathbf Y_a{}^{bc}&=
     -\mathbf k_{b}\cdot \mathbf k_{c} \frac{-i\mathbf k_c}{k_c^2} S^{(n)}(k_c)\notag \\
    & \times (2\pi)^3\delta^{(3)}_{\rm D}(\mathbf k_a-\mathbf k_b-\mathbf k_c),\\
    \mathbf Y_a{}^{bcd}& = (\mathbf k_{b}\cdot \mathbf k_{d})(  \mathbf k_{c} \cdot  \mathbf   k_{d} ) \frac{-i\mathbf k_d}{k_d^2} S^{(n)}(k_d)\notag \\
    &\times (2\pi)^3\delta^{(3)}_{\rm D}(\mathbf k_a-\mathbf k_b-\mathbf k_c-\mathbf k_d).
\end{align}
Then, expanding with respect to $\phi$ using Eq.~\eqref{deltaNLexp}, we get
\begin{align}
	\mathbf t_a &= \mathbf Y_a{}^s
X_{s}{}^p A_p{}^b \phi_b 
	+
	\frac{1}{2}\mathbf Y_a{}^s X_{s}{}^p A_p{}^{bc}\phi_b\phi_c  
	\notag \\
	&	
	+ \frac{1}{2}\mathbf Y_a{}^s X_s{}^{pq}
	A_p{}^b  A_q{}^c \phi_b\phi_c
		\notag \\
	&+
    \frac12\mathbf Y_a{}^{ts}X_{s}{}^p 
    A_t{}^{(b}  A_p{}^{c)} \phi_b \phi_c 
    \notag \\
	&
	+
	\frac{1}{3!}\mathbf Y_a{}^s X_{s}{}^pA_p{}^{bcd} \phi_b  \phi_c  \phi_d
		\notag \\
	&	
	+\frac{1}{3!}\mathbf Y_a{}^s  X_s{}^{pq}A_p{}^{(bc} A_q{}^{d)}  \phi_b  \phi_c  \phi_d
			\notag \\
	&	
    +\frac{1}{3!}
    \mathbf Y_a{}^{ts}  
	 X_{s}{}^p A_p{}^{(cd} A_t{}^{b)} \phi_b \phi_c\phi_d
    \notag \\
    &
   + \frac{1}{3!} \mathbf Y_a{}^{ts} X_{s}{}^p  A_t{}^{(bc}  A_p{}^{d)} \phi_b\phi_c \phi_d
   		\notag \\
	&	
+\frac{1}{3!}
    \mathbf Y_a{}^{ts}  X_s{}^{pq}
	A_p{}^{(c}  A_q{}^d A_t{}^{b)} \phi_b \phi_c\phi_d
    	\notag \\
	&	
  +
	\frac{1}{3!}\mathbf Y_a{}^sX_s{}^{pqr} A_p{}^bA_q{}^cA_r{}^d  \phi_b  \phi_c  \phi_d
		\notag \\
	&	
+
    \frac{1}{3!}\mathbf Y_a{}^{tus} X_{s}{}^p A_t{}^{(b}A_u{}^cA_p{}^{d)} \phi_b \phi_c   \phi_d,\label{shiftvecn}
\end{align}
where the parenthesis implies that we take all possible permutations without duplication.
Thus, we obtained the perturbative expansion of the shift vector in Fourier space.
We are ready to shift the density field using Eq.~\eqref{shiftvecn}.

\subsubsection{Next step displacement}

Finally, we update the displacement field by following Eq.~\eqref{psin}.
Ignoring the curl component, Eq.~\eqref{psinexetk} yields
\begin{align}
	\phi^{(n+1)} = \phi^{(n)} + \frac{-i\mathbf k}{k^2}\cdot \mathbf t^{(n)}.
\end{align}
For all $\mathbf Y$ we introduce
\begin{align}
	Y_{a}{}^{b\cdots} \equiv  \frac{-i\mathbf k_a}{k_a^2}\cdot \mathbf Y_{a}{}^{b\cdots}.
\end{align}
Then we find the following recurrence relations:
\begin{align}
	A^{(n+1)}_a{}^b &= A^{(n)}_a{}^b
	+ Y_a{}^s X_{s}{}^p A^{(n)}_p{}^b ,\label{rec1_}
	\\
	A^{(n+1)}_a{}^{bc} &=A^{(n)}_a{}^{bc} 
	+
	Y_a{}^s X_{s}{}^p A^{(n)}_p{}^{bc}
		\notag \\
	&
	+
	 Y_a{}^s X_s{}^{pq}
	A^{(n)}_p{}^b  A^{(n)}_q{}^c 
	\notag \\
	&
	+Y_a{}^{ts}X_{s}{}^p 
    A^{(n)}_t{}^{(b}  A^{(n)}_p{}^{c)}  ,\label{rec2_2}
	 \\
    A^{(n+1)}_a{}^{bcd} &=A^{(n)}_a{}^{bcd}
    +Y_a{}^s X_{s}{}^pA^{(n)}_p{}^{bcd}
	\notag \\
	&
	+Y_a{}^s  X_s{}^{pq}A^{(n)}_p{}^{(bc} A^{(n)}_q{}^{d)} 
		\notag \\
	&
    +
    Y_a{}^{ts}  
	 X_{s}{}^p A^{(n)}_p{}^{(cd} A^{(n)}_t{}^{b)} 
    \notag \\
    &
   + Y_a{}^{ts} X_{s}{}^p  A^{(n)}_t{}^{(bc}  A^{(n)}_p{}^{d)} 
       \notag \\
    &
   	+
    Y_a{}^{ts}  X_s{}^{pq}
	A^{(n)}_p{}^{(c}  A^{(n)}_q{}^d A^{(n)}_t{}^{b)} 
    \notag \\
    &
      +Y_a{}^sX_s{}^{pqr} A^{(n)}_p{}^bA^{(n)}_q{}^cA^{(n)}_r{}^d  
          \notag \\
    &
+Y_a{}^{tus} X_{s}{}^p A^{(n)}_t{}^{(b}A^{(n)}_u{}^cA^{(n)}_p{}^{d)},\label{rec3_3}
\end{align}
where the parentheses mean that we take the possible permutations to symmetrize the indices.
Thus, we obtain the recurrence relation for the expansion kernels defined in Eq.~\eqref{eq27}.
Even though the recurrence relations misleadingly appear complicated, they only depend on the smoothing scales and are independent of cosmology. We can therefore precompute them.
Using the above recurrence relation and the initial conditions~\eqref{A1_0} and \eqref{A23_0}, we can compute the $n$-th step displacement field systematically.
Explicit forms of Eqs. \eqref{rec1_} to \eqref{rec3_3} without index are presented in Appendix~\ref{woind}.
Eq.~\eqref{rec1_} straightforwardly reduces to Eq.~\eqref{Eq.A1}, and we get 
\begin{align}
    A^{(n)} = (1 - S^{(n-1)})(1 - S^{(n-2)})\cdots (1 - S^{(0)}).
\end{align}
Thus, $A^{(n)}\to 0$ for $n\to \infty$ and the displacement fields in the later steps converge to zero as approaching the estimated Lagrangian position.
The second and third-order kernels are more complicated, but we will see that we only use a few specific momentum configurations in the following calculations.

\subsection{Post iterative reconstruction power spectrum}

From Eqs~\eqref{def:rec} and \eqref{abs28}, we get the reconstructed displacement potential
\begin{align}
	\phi^{(n)}_{{\rm rec},a} =& \bar A^{(n)}_a{}^b \phi_b 
	-
	\frac{1}{2}A^{(n)}_a{}^{bc}\phi_b\phi_c  -
	\frac{1}{3!}A^{(n)}_a{}^{bcd}\phi_b\phi_c\phi_d,
\end{align}
where $\bar A^{(n)}_a{}^b = \delta_a{}^b - A^{(n)}_a{}^b $, with $\delta_a{}^b\equiv (2\pi)^3\delta^{(3)}_{\rm D}(\mathbf k_a-\mathbf k_b)$.
The 1-loop power spectrum of the estimator is 
\begin{align}
	\langle \phi^{(n)}_{{\rm rec,}a}  \phi^{(n)}_{{\rm rec,}e}\rangle 
	&= \bar A^{(n)}_a{}^b\bar A^{(n)}_e{}^f \langle \phi_b 
	 \phi_f \rangle
	\notag \\
	&-
	\bar A^{(n)}_a{}^b  A^{(n)}_e{}^{fg}\langle \phi_b\phi_f\phi_g \rangle  \notag \\
	&-\frac{1}{3}
	\bar A^{(n)}_a{}^b A^{(n)}_e{}^{fgh}\langle \phi_b\phi_f\phi_g\phi_h \rangle
	\notag \\
	&+
	\frac{1}{4}A^{(n)}_a{}^{bc}
	A^{(n)}_e{}^{fg}\langle \phi_b\phi_c\phi_f\phi_g \rangle.\label{eq64}
\end{align}
We compute Eq.~\eqref{eq64} by expanding $\phi$ with respect to $\delta_{\rm L}$:
\begin{align}
	\phi_a =& L_a{}^b\delta_{{\rm L},b} +\frac{1}{2}L_a{}^{bc}\delta_{{\rm L},b}
	\delta_{{\rm L},c}
	+\frac{1}{3!}L_a{}^{bcd}\delta_{{\rm L},b}
	\delta_{{\rm L},c}
	\delta_{{\rm L},d}\,,\label{phiex}
\end{align}
where the Lagrangian kernels are written as~\cite{Matsubara:2007wj,Baldauf:2015tla}
\begin{align}
	 L_a{}^b&= \frac{1}{k_a^2}(2\pi)^3 \delta^{(3)}_{\rm D}(\mathbf k_a-\mathbf k_b),\label{L1}\\
	 L_a{}^{bc}&=\frac{3}{7}\frac{1}{k_a^2}\frac{(\mathbf k_b\times \mathbf k_c)^2}{|\mathbf k_b|^2|\mathbf k_c|^2} (2\pi)^3  \delta^{(3)}_{\rm D}(\mathbf k_a-\mathbf k_b-\mathbf k_c),\label{L2}\\
	 L_a{}^{bcd}&=\left[	-\frac{1}{3}\frac{1}{k_a^2}\frac{[\mathbf k_d\cdot (\mathbf k_b \times \mathbf k_c)]^2}{|\mathbf k_d|^2|\mathbf k_b|^2|\mathbf k_c|^2} \right. \notag \\
	 &\left. +\frac{5}{21}\frac{1}{k_a^2}\frac{ [\mathbf k_b\times (\mathbf k_c+\mathbf k_d)]^2}{|\mathbf k_b|^2(\mathbf k_c+\mathbf k_d)^2}\frac{(\mathbf k_c\times \mathbf k_d)^2}{|\mathbf k_c|^2|\mathbf k_d|^2}+2~{\rm perms}\right]
	 \notag \\
	 &\times (2\pi)^3\delta^{(3)}_{\rm D}(\mathbf k_a-\mathbf k_b-\mathbf k_c-\mathbf k_d)
.\label{L3}
\end{align}

Then the 1LPT auto-power spectrum of $\phi^{(n)}_{\rm rec}$ and its cross-power spectrum with $\delta_{\rm L}$ are straightforwardly obtained as
\begin{align}
	 P^{{\rm 1LPT}}_{\phi^{(n)}_{\rm rec}} = A^{(n)2}P^{\rm 1LPT}_{\phi},~ P^{{\rm 1LPT}}_{\delta_{\rm L} \phi^{(n)}_{\rm rec}} = A^{(n)}P^{\rm 1LPT}_{\phi}.\label{rec:zel:pow}
\end{align}
Then the 3LPT postreconstruction power spectrum is given as follows: 
\begin{align}
	&P^{{\rm 3LPT}}_{\phi^{(n)}_{\rm rec}}=	 
	P^{{\rm 3LPT}}_{\phi^{(n)}_{\rm rec}11}
	+
	P^{{\rm 3LPT}}_{\phi^{(n)}_{\rm rec}13}
	+
	P^{{\rm 3LPT}}_{\phi^{(n)}_{\rm rec}22} ,\label{1-loop:power}
\end{align}	
where we defined
\begin{align}
    P^{{\rm 3LPT}}_{\phi^{(n)}_{\rm rec}11} = & \bar A^{(n)}(k)^2 P^{\rm 3LPT}_\phi(k), \\
    P^{{\rm 3LPT}}_{\phi^{(n)}_{\rm rec}22} =	& - \int dx  d\mu\frac{P_{\rm L}\left(ky\right)P_{\rm L}(kx)}{4\pi^2k}\notag \\
	 \times &\left(\frac{3}{7y^4} (1-\mu^2)\bar A^{(n)}(k) \alpha^{(n)} - \frac{\alpha^{(n)2} }{2x^2y^4} \right) ,
	 \\
P^{{\rm 3LPT}}_{\phi^{(n)}_{\rm rec}13}=	& -
	\bar A^{(n)}(k)P_{\rm L}(k)\int dx  d\mu \frac{P_{\rm L}(kx)}{4\pi^2k} \notag \\
	\times &
	\left( 
	\frac{\beta^{(n)}}{x^2} +    \frac{6}{7y^2}(1-\mu^2) \alpha^{(n)}  \right) ,
\end{align}
where $k'=kx$ is the loop momentum, and $\mu$ is the cosine defined for $\mathbf k$ and the loop momentum, $y\equiv \sqrt{1-2x\mu+x^2}$, and $\alpha$, $\beta$ and $\gamma$ are given as 
\begin{align}
	\alpha^{(n)}(k,k',\mu) &\equiv A^{(n)}(\mathbf k-\mathbf k',\mathbf k')/k^2,\\
	\beta^{(n)}(k,k',\mu) &\equiv A^{(n)}(\mathbf k,\mathbf k',-\mathbf k')/k^4, \\
	\gamma^{(n)}(k,k',\mu) &\equiv  A^{(n)}(\mathbf k,-\mathbf k')/k^2\, .
\end{align}
\eqref{A2} and \eqref{rec3_3_s} can be reduced to 
\begin{align}
	\alpha^{(n+1)} &=\alpha^{(n)}
	-S^{(n)}(k)\alpha^{(n)} \notag \\
	&-
	x\mu( 1-x\mu) 
	S^{(n)}(k)A^{(n)}(kx) A^{(n)}(ky) 
	\notag \\
	&- x( x- \mu  ) (1-x\mu)   S^{(n)}(ky)
	A^{(n)}(kx)  A^{(n)}(ky) \notag \\
	&
	- x^2\mu( x- \mu  )  S^{(n)}(kx) A^{(n)}(kx)  A^{(n)}(ky),\label{recalpha}
\\
	\beta^{(n+1)} &= \beta^{(n)}
    -S^{(n)}(k)\beta^{(n)}
     \notag \\
    &
      +2 x^2\mu^2  S^{(n)}(k)  A^{(n)}(k)A^{(n)}(kx)^2
	\notag \\
	&
	+  2x^4\mu^2  S^{(n)}(kx) A^{(n)}(k)A^{(n)}(kx)^2
	\notag \\
    &
	-2x\mu
	(1-x\mu) S^{(n)}(k) A^{(n)}(kx) \gamma^{(n)}  
	\notag \\
    &
   - 2x^2\mu(x-\mu) S^{(n)}(kx) A^{(n)}(kx)\gamma^{(n)}
   \notag \\
	&   -
	2x(x-\mu )(1-x\mu) S^{(n)}(ky)  A^{(n)}(kx)\gamma^{(n)} 
   	\notag \\
	&-
    2\frac{ x^2(x-\mu)^2 (1-x\mu)^2  }{y^2} S^{(n)}(ky)
	A^{(n)}(k)A^{(n)}(kx)^2 
  ,\label{recbeta}
\\
	\gamma^{(n+1)} &=\gamma^{(n)} 
	-S^{(n)}(ky)\gamma^{(n)} \notag \\
	&- 
	\frac{x(x-\mu) (1-x\mu)  }{y^2} 
	S^{(n)}(ky) A^{(n)}(k) A^{(n)}(kx) 
	\notag \\
	&-\frac{x^2\mu}{y^2}  (x-\mu) S^{(n)}(kx) A^{(n)}(k)  A^{(n)}(kx) \notag \\
	&-\frac{x\mu}{y^2} (1-x\mu) S^{(n)}(k) 
    A^{(n)}(k)  A^{(n)}(kx)
		.\label{recgamma}
\end{align}
Note that Eq.~\eqref{recbeta} is valid under the premise that we integrate $\beta^{(n)}$ with $\mu^2$ from $\mu=-1$ to 1, and its original expression was the symmetric version of Eq.~\eqref{recbeta} with respect to $\mu \to -\mu$.  
Similarly, we get the 3LPT cross-power spectrum
\begin{align}
	&P^{{\rm 3LPT}}_{\delta_{\rm L}\phi^{(n)}_{\rm rec}} 
	=\bar A^{(n)}(k) \left[P^{\rm 1LPT}_{\phi}+\frac{P_{\phi13}}{2} \right] +\frac{1}{2}P^{{\rm 3LPT}}_{\phi^{(n)}_{\rm rec}13}
 .\label{1-loop:prop}
\end{align}

To summarize, although we have derived a strenuous set of equations so far, we have straightforwardly followed the steps in Sec.~\ref{moderec}, expanding the equations up to third order in the perturbations.
Readers who are only interested in the main result can refer to Eqs.~\eqref{1-loop:power} and \eqref{1-loop:prop} and the associated recurrence relations.

\subsection{Post iterative reconstruction propagator and cross-correlation coefficient}

The propagator is defined as the cross-correlation function between a late time field $X$ and the linear matter perturbation $\delta_{\rm L}$ normalized by the linear spectrum:
\begin{align}
	C_{X}\equiv \frac{P_{\delta_{\rm L} X}}{P_{\rm L}}.\label{cc}
\end{align}
While not directly observable in real data, this quantity is a useful metric for density field reconstruction techniques, since it quantifies how much of the initial linear perturbation has been recovered from the (measured) late time density as a function of the scale $k$.
One also uses the cross-correlation coefficient 
\begin{align}
	r_{XY}\equiv \frac{P_{XY}}{\sqrt{P_{XX}P_{YY}}},\label{rr}
\end{align}
to quantify the reconstruction efficiency.
The difference between Eqs~\eqref{cc} and \eqref{rr} is the normalization, and the two represent different information as explained below.
Consider $X=\delta_{\rm L}$ and $Y=k^2\phi$.
The 1-loop cross-correlation coefficient is
\begin{align}
	r^{\rm 3LPT}_{\delta_{\rm L} k^2\phi} = \frac{P_{\phi11}+\frac{1}{2}P_{\phi13}}{\sqrt{P_{\phi11}(P_{\phi11}+P_{\phi13}+P_{\phi22})}}. \label{eqcrossco}
\end{align}
Then, $P_{\phi 11}\gg|P_{\phi13}|, P_{\phi22}$ yields
\begin{align}
	r^{\rm 3LPT}_{\delta_{\rm L} k^2\phi} \sim 1 -\frac{1}{2}\frac{P_{\phi22}}{P_{\phi11}}+\mathcal O\left(\frac{P^2_{\phi22}}{P^2_{\phi11}},\frac{P^2_{\phi13}}{P^2_{\phi11}}\right).
\end{align}
Thus, $P_{\phi13}$ vanishes at the leading order even if $|P_{\phi13}|\gtrapprox P_{\phi22}$.
Hence, the cross-correlation coefficient is a measure of $P_{\phi22}$, i.e., the mode coupling.
In simulations or observations, noises also appear in the same way.
On the other hand, the propagator is given as
\begin{align}
	C^{\rm 3LPT}_{k^2\phi} = 1 + \frac{1}{2}\frac{P_{\phi13}}{P_{\phi11}},
\end{align}
being a measure of $P_{\phi13}$, i.e., the shift.
In the following section, we will present both the propagator and the cross-correlation coefficient to quantify how well the LPT modeling works.

\subsection{Theory perspective of the iterative reconstruction as a regularization method}

As we already mentioned, the prereconstruction 3LPT is known to suffer the UV mistake when modeling the nonlinear displacement field.
In field theory, we have a similar situation.
We often encounter superficial divergences while actual physical quantities are finite.
Such divergences appear only at intermediate stages of calculation, which can be canceled after regularization.
The prereconstruction 3LPT loop integral UV sensitivity is a similar issue, so it is a natural question to ask how we regularize the loop integral to compute the observable properly.

Assuming that the iterative reconstruction correctly returns the true displacement field in the end, we may think of the procedure as another way of calculating the displacement field from the density field.
The reconstructed displacement field is not necessarily the same as the naively defined displacement field in Eq.~\eqref{psidef}, and there is no prior reason to believe that the latter is the correct one as the 3LPT prereconstruction includes the UV mistake.
Indeed, the iterative reconstruction can be a more clever alternative to estimate the Lagrangian displacement since we avoid directly computing the nonlinearity.
That is, we smooth the nonlinear part of $\delta_{\rm NL}$ step by step, and at every step, the linear approximation is quite good for the continuity equation.
Therefore, we may think of the iterative reconstruction to correct the UV mistake, i.e., it can work as a regularization of the UV-sensitive loop integral.

On the analogy of field theory, we may regard the displacement field in Eq.~\eqref{psidef} as a ``bare'' quantity, which can be divergent at the intermediate stage of the calculation.
Then the reconstructed displacement field $\phi^{(n)}_{\rm rec}$ is ``physical'' because the reconstruction procedures are based on the observed energy-momentum tensor, i.e., the mass density.
The iterative procedure naturally introduces the counter term as  Eq.~\eqref{1-loop:power} can be recast into
\begin{align}
    P^{\rm 3LPT}_{\phi^{(n)}_{\rm rec}} &= P^{\rm 3LPT}_{\phi} - P^{{\rm ct}(n)},
\end{align}
where we introduced
\begin{align}
        P^{\rm ct(n)} \equiv & (2A^{(n)}-A^{(n)2})P^{\rm 3LPT}_{\phi} - P^{{\rm 3LPT}}_{\phi^{(n)}_{\rm rec}13}
	-
	P^{{\rm 3LPT}}_{\phi^{(n)}_{\rm rec}22}.
\end{align}
Below, we will see that $P^{\rm 3LPT}_{\phi^{(n)}_{\rm rec}}$ agrees with the true displacement field up to $k{\rm Mpc}/h<0.2$; therefore, this is a physically well-motivated way to introduce a counter term.
As a caveat, a simple sharp cut-off in the integral \eqref{P13} violates the momentum space transnational symmetry and cannot reproduce the true displacement field. 
The regularization would work up to some $n$ corresponding to the UV cut-off scale of the theory described by Eq.~\eqref{eom1}.
In this scheme, we progressively include the modes from IR to UV and find the scale where the UV modes are no more reliable.
Indeed, the true displacement field has a peak at $k{\rm Mpc}/h\sim 0.2$ and is quickly damped for higher $k$ in Fig.~\ref{fig00}.
Therefore, the linear displacement is overestimated for $k{\rm Mpc}/h> 0.2$ in the perturbation theory, and so is the loop integral.

\section{Comparison with simulations}
\label{sec5}

In this section, we compare our LPT modeling with simulations.
Let us first present the simulation results and see if the iterative procedures are successfully modeled.

\subsection{Simulation of post reconstruction field}\label{simulation:sec}

\begin{figure}
  \includegraphics[width=\linewidth]{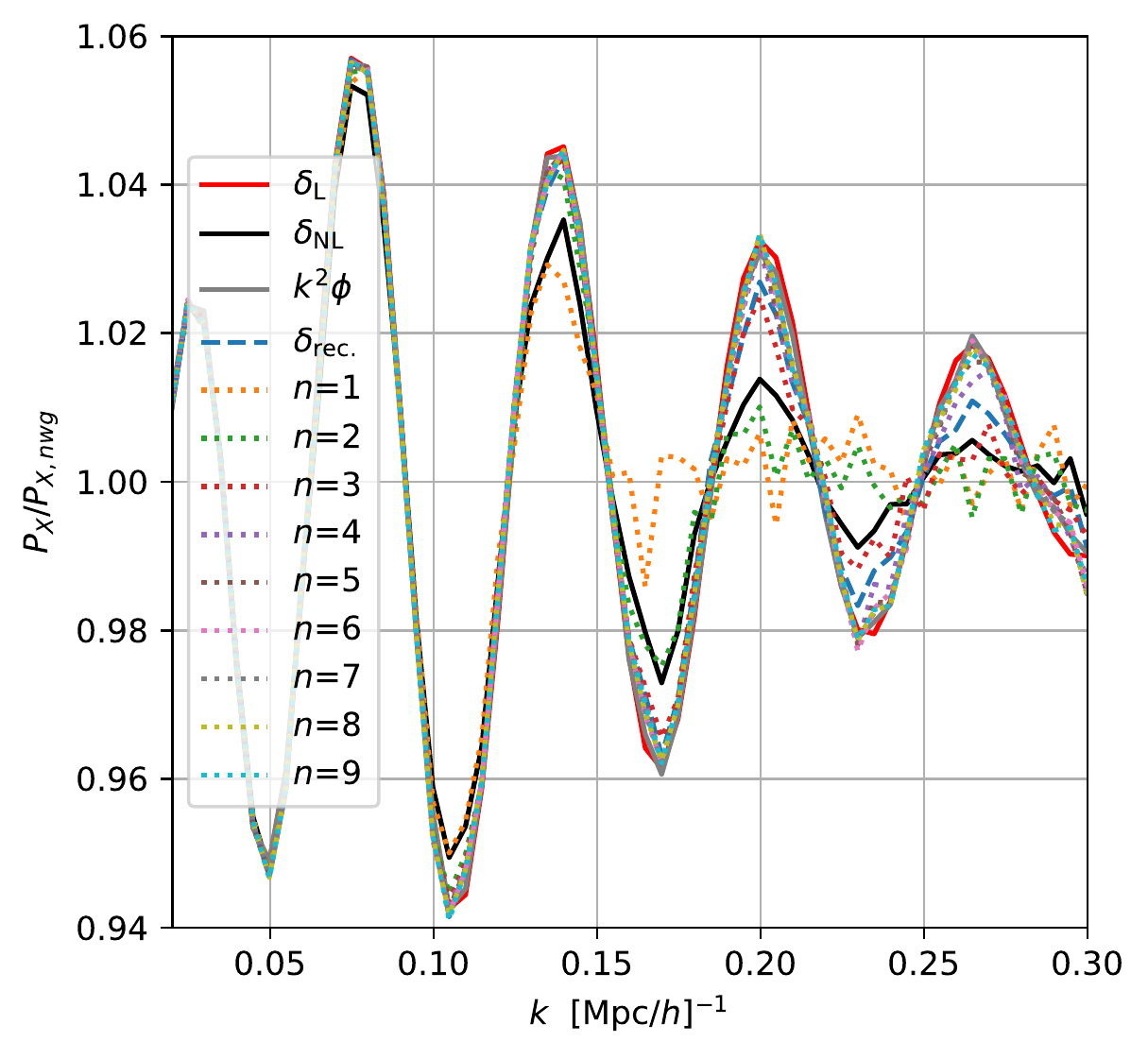}
  \caption{power spectrum divided by its matched no-wiggle one. The data set is L1500. The no-wiggle spectra are defined for each simulation and thus any nonlinear and numerical effect on the broadband shape are removed. $\delta_{\rm rec}$ is the standard reconstruction with $R=20$Mpc/$h$. }
  \label{fig2}
\end{figure}

We choose the initial ($n=0$) smoothing scale of $20 {\rm Mpc}/h$ to reconstruct L500 and subL500, and we reduce the smoothing scale incrementally by $\epsilon=1/\sqrt{2}$ until we reach $n=9$.
In Fig.~\ref{fig2}, we reproduced Fig.~5 in Ref.~\cite{Schmittfull:2017uhh}, adding displacement fields.
We plot the nonlinear fields and postreconstruction fields normalized by the corresponding no-wiggle spectra. The no-wiggle spectra are defined for each simulation, calculated from the paired no-wiggle simulation in Sec.\ref{simsec}, and thus the nonlinear and numerical effect on the broadband shape is removed. Hence, we can single out the degradation effect in the BAO.
As we see in the plot, the initial density, nonlinear displacement, and the iterative reconstruction estimators agree well below $k{\rm Mpc}/h<0.3$. We overplot the standard reconstruction $\delta_{\rm rec}$ with the smoothing scale of $20 {\rm Mpc}/h$ (blue lines); its reconstructed BAO feature is less distinct than $n>7$ mainly because the smoothing scale is smaller than the effective smoothing scale of the iterative reconstructions~\cite{Seo:2021}.

\begin{figure}
  \includegraphics[width=\linewidth]{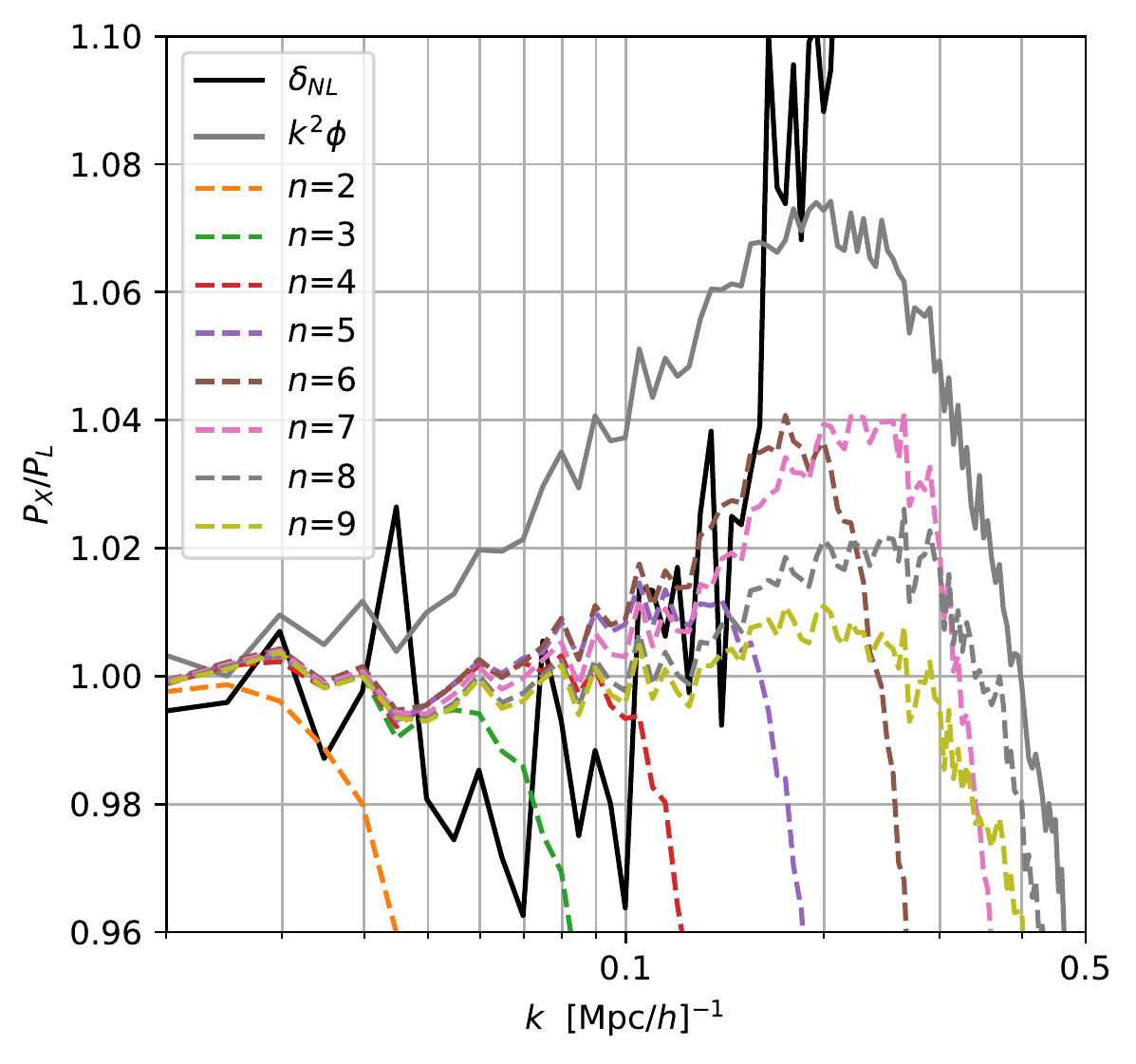}
  \caption{Power spectra normalized by the linear matter power spectrum for L500. One can see the effect of reconstruction on the full shape.}
  \label{fig22}
\end{figure}

\begin{figure}
  \includegraphics[width=\linewidth]{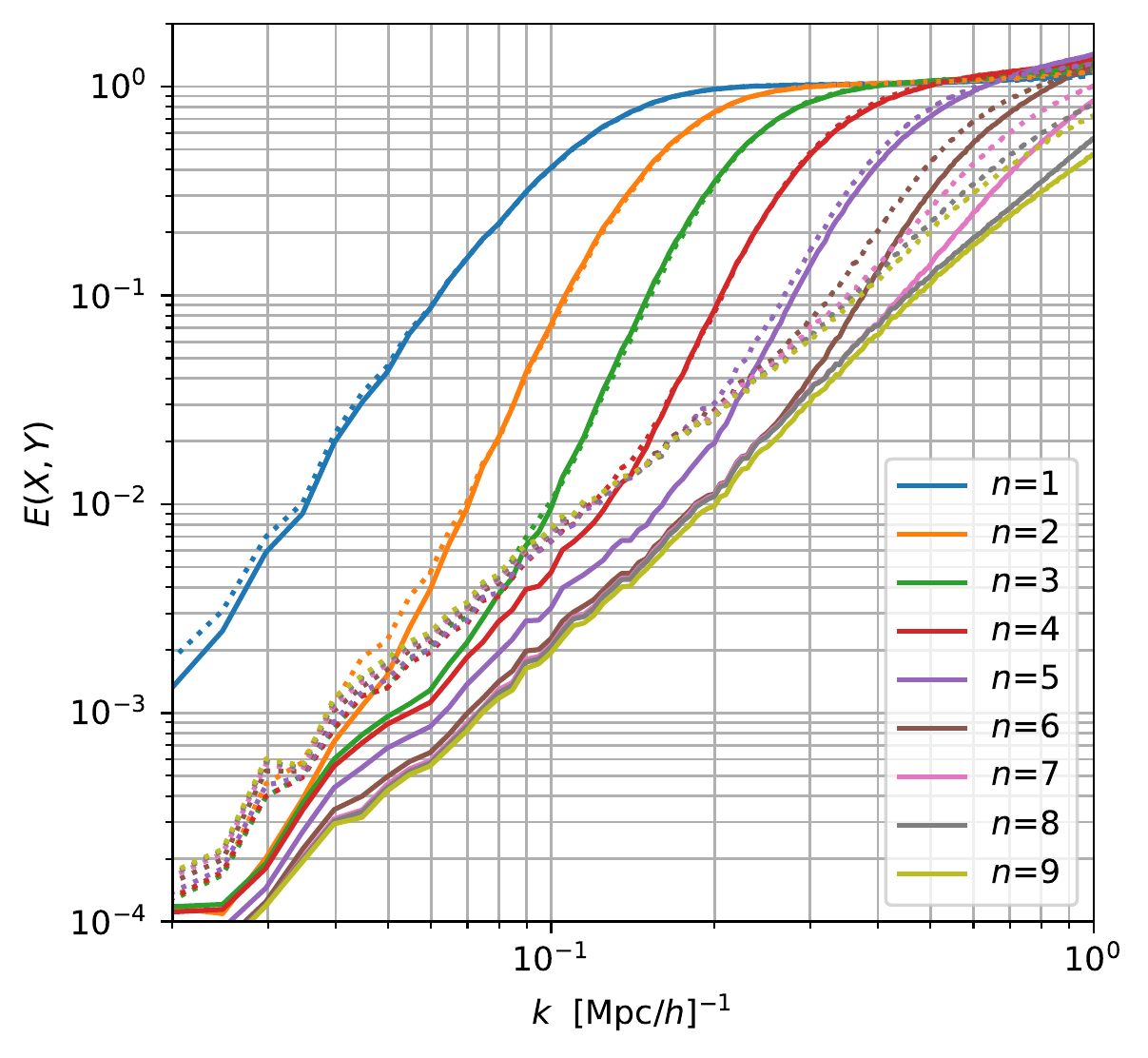}
  \caption{The error power spectrum $E(k^2\phi^{(n)}_{\rm rec},Y)$ for $Y=k^2\phi$~(solid) and $Y=\delta_{\rm L}$ (dotted). A lower error power spectrum implies a better agreement. The postreconstruction field is ``closer'' to the displacement field than the linear field. Iterative reconstruction is convergent around $n=6$ and the error is at most 3\% for $k$Mpc/$h<$0.3.}
  \label{fig222}
\end{figure}

To see the rest of the nonlinear effects, we plotted the nonlinear power spectrum normalized by the linear one in Fig.~\ref{fig22}.
Our $n=9$ visually agrees with Fig.14 in Ref.~\cite{Schmittfull:2017uhh}. The higher noise in the $\delta_{\rm NL}$ power spectrum in the figure, compared to the power spectra of displacement field tracers, reflects its poor cross-correlation with the initial field. The iterative reconstruction appears to converge to the displacement field rather than to the linear power spectrum as $n$ approaches seven and then approaches closer to the linear shape as $n$ increases.
To better quantify how close the reconstructed field is to the nonlinear displacement field or to the linear field, 
we introduce the error power spectrum of $X$ and $Y$ as
\begin{align}
    E(X,Y) = \frac{P_{X-Y}}{P_Y},\label{erdef}
\end{align}
where $P_{X-Y}$ and $P_Y$ are the auto-power spectrum of $X-Y$ and $Y$, respectively. 
Then we show $E(k^2\phi^{(n)}_{\rm rec},k^2\phi)$ and $E(k^2\phi^{(n)}_{\rm rec},\delta_{\rm L})$ in Fig.~\ref{fig222}.
We can see $E(k^2\phi^{(n)}_{\rm rec},k^2\phi)<E(k^2\phi^{(n)}_{\rm rec},\delta_{\rm L})$, and thus the postreconstruction field is relatively ``closer'' to the displacement field than the linear field as iteration increases in terms of the error power spectrum.
Fig.~\ref{fig222} shows that the iterative reconstruction works as expected, i.e., gradually reduces the error power spectrum progressively to a smaller scale; most of the convergence is reached around $n=7$, and there is a slight improvement for $n>7$.

 Figs \ref{1-loopres1}, \ref{1-loopres1.1} and \ref{1-loopres3.1} show the power spectrum, the propagator, and the cross-correlation coefficient in each step of iteration. At $n=6$ and 7, we indeed see a slight positive shift term, similar to that of the displacement (grey), but such feature is very weak and disappears after $n>7$. In terms of the cross-correlation coefficient, again, the reconstructed field converges to the displacement field up to $k {\rm Mpc}/h \sim 0.2$  for $n>5$. 
 
 In summary, the iterative reconstruction converges to the displacement field in mode coupling terms, but it deviates in terms of the shift term. For the shift term, the reconstructed field appears to converge to the linear field ($k {\rm Mpc}/h \sim 0.2$).

\subsection{LPT modeling and simulations}

\begin{figure*}
  \includegraphics[width=\linewidth]{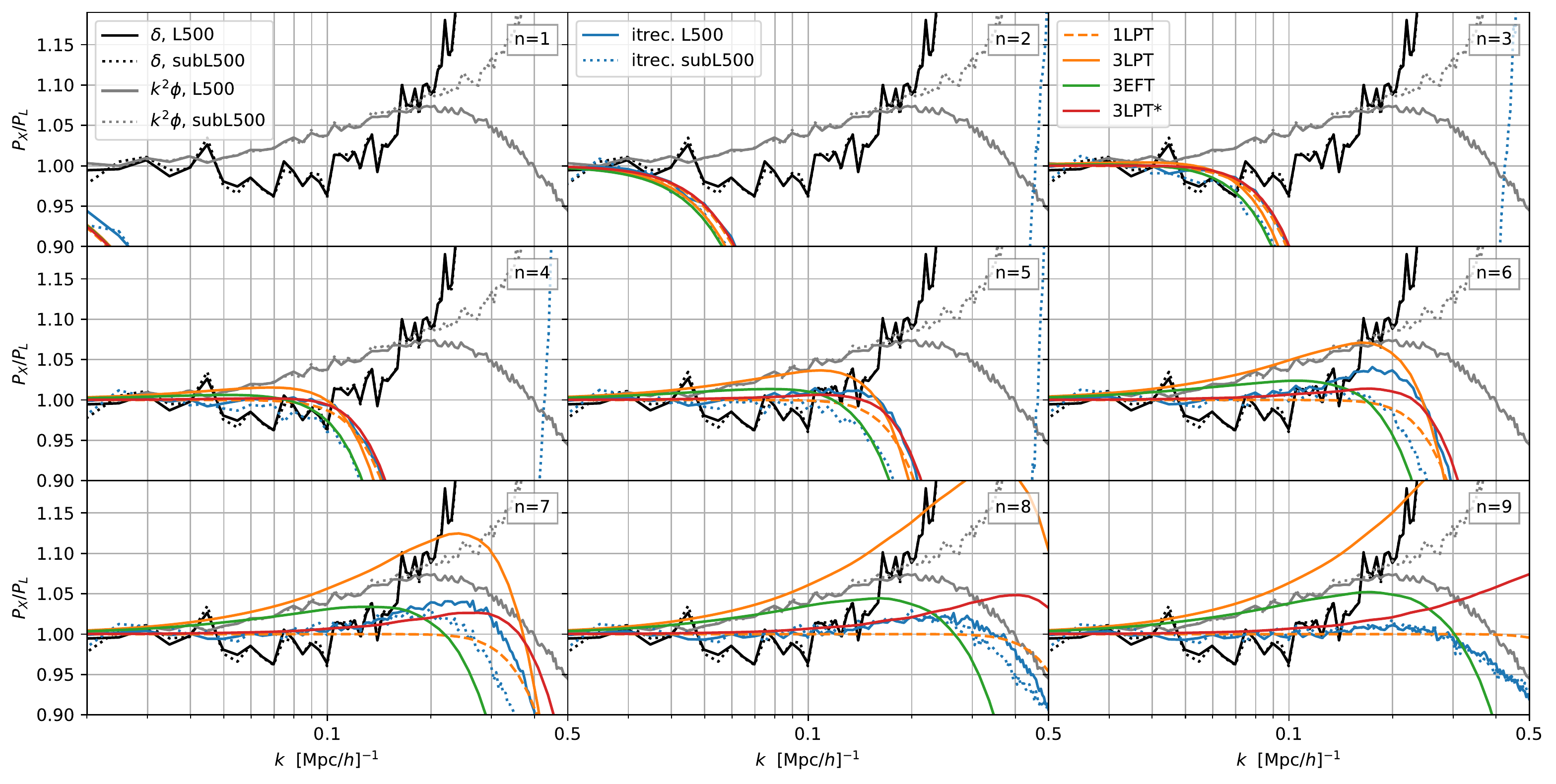}
  \caption{The comparison between the simulated postreconstruction power spectra~(blue), simulated true displacement~(gray) and the 3LPT~\eqref{1-loop:power}, 1LPT~\eqref{rec:zel:pow}, 3EFT~\eqref{3EFT} and 3LPT*~\eqref{1-loop:power22} models.
  Solid and doted curves for the simulation are 4\% and 0.15\% subsampling, respectively.
  Our 3LPT prediction agrees with the true displacement field rather than the simulated postreconstruction displacement up to $n=6$.
  3LPT* is comparable to the simulated post reconstruction displacement field within 1\% accuracy.
  The EFT slowly converges to the true displacement field. 
  }
  \label{1-loopres1}
   \includegraphics[width=\linewidth]{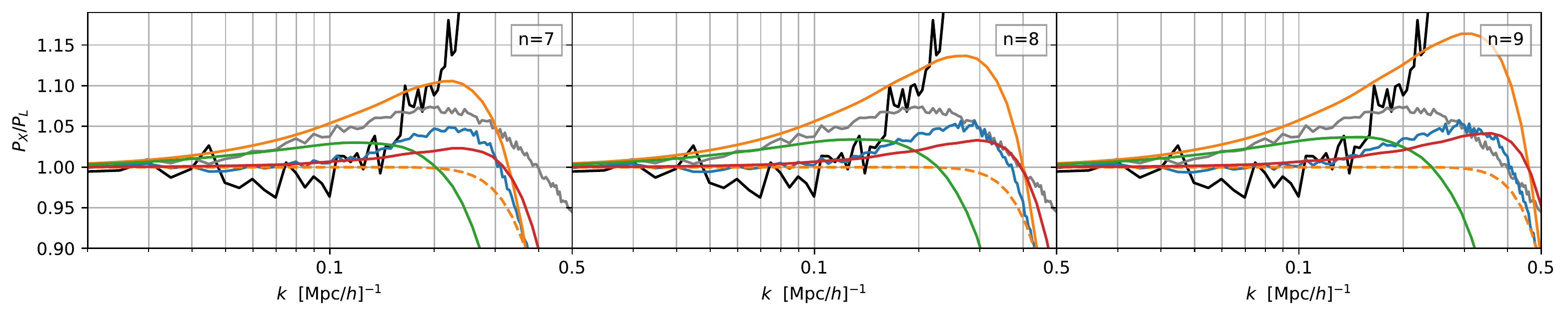}
  \caption{The power spectrum in Fig.~\ref{1-loopres1} for the minimum smoothing scale $R_{\rm min}=3.5{\rm Mpc}/h$ Plots are the same with Fig.~\ref{1-loopres1} up to $n=6$ and are almost fixed at $n=7$. 
  }
  \label{1-loopres2}
 \end{figure*}

\begin{figure*}
   \includegraphics[width=\linewidth]{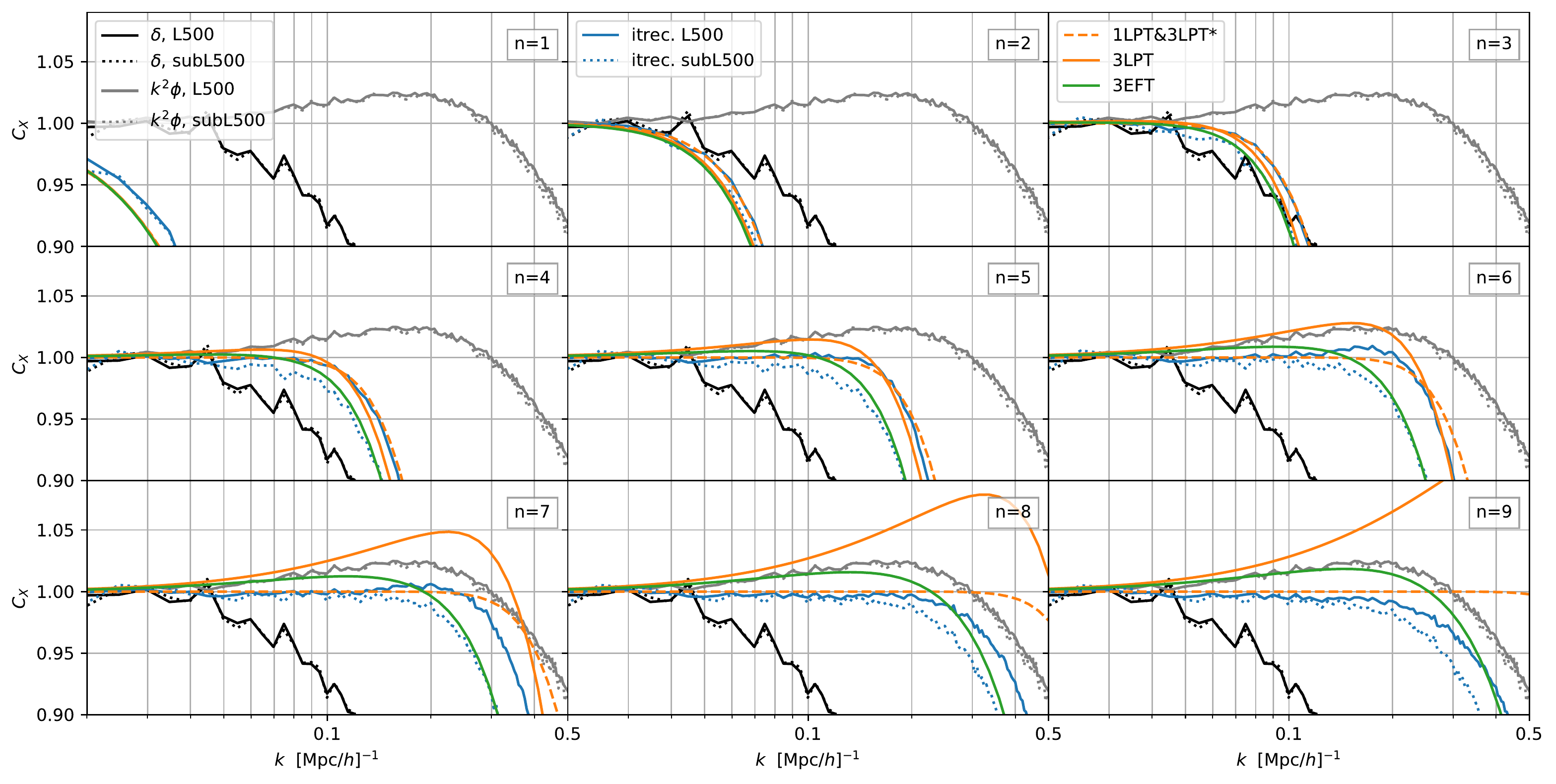}
  \caption{The propagator counterparts of Fig.~\ref{1-loopres1}. The simulated displacement field, 3LPT, 3EFT show the shift term, but the effect is very small for the simulated postreconstruction.}
  \label{1-loopres1.1}

  \includegraphics[width=\linewidth]{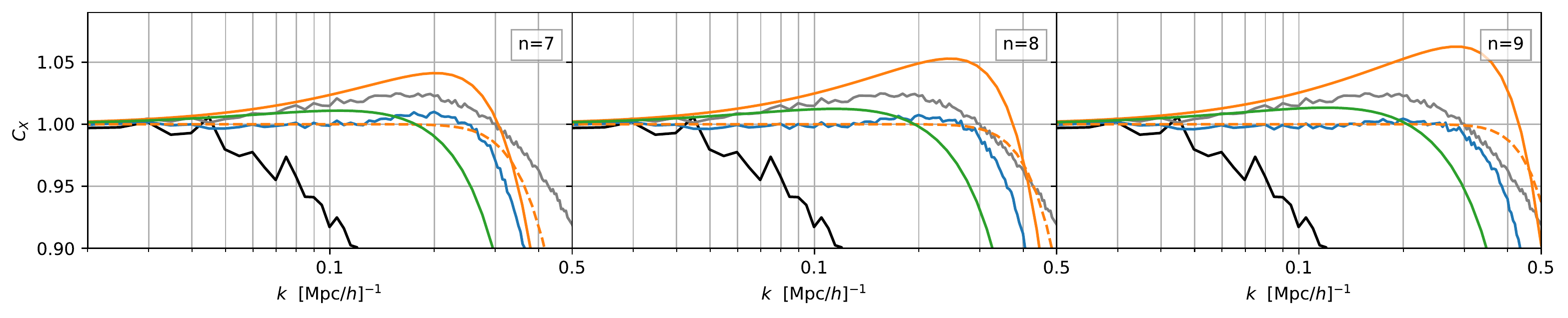}
  \caption{The propagator in Fig.~\ref{1-loopres1.1} with the minimum smoothing scale $R_{\rm min}\sim 3.5$Mpc/$h$.
  Plots are the same with Fig.~\ref{1-loopres1.1} up to $n=6$.
  }
  \label{1-loopres2.1}
\end{figure*}

\begin{figure*}
   \includegraphics[width=\linewidth]{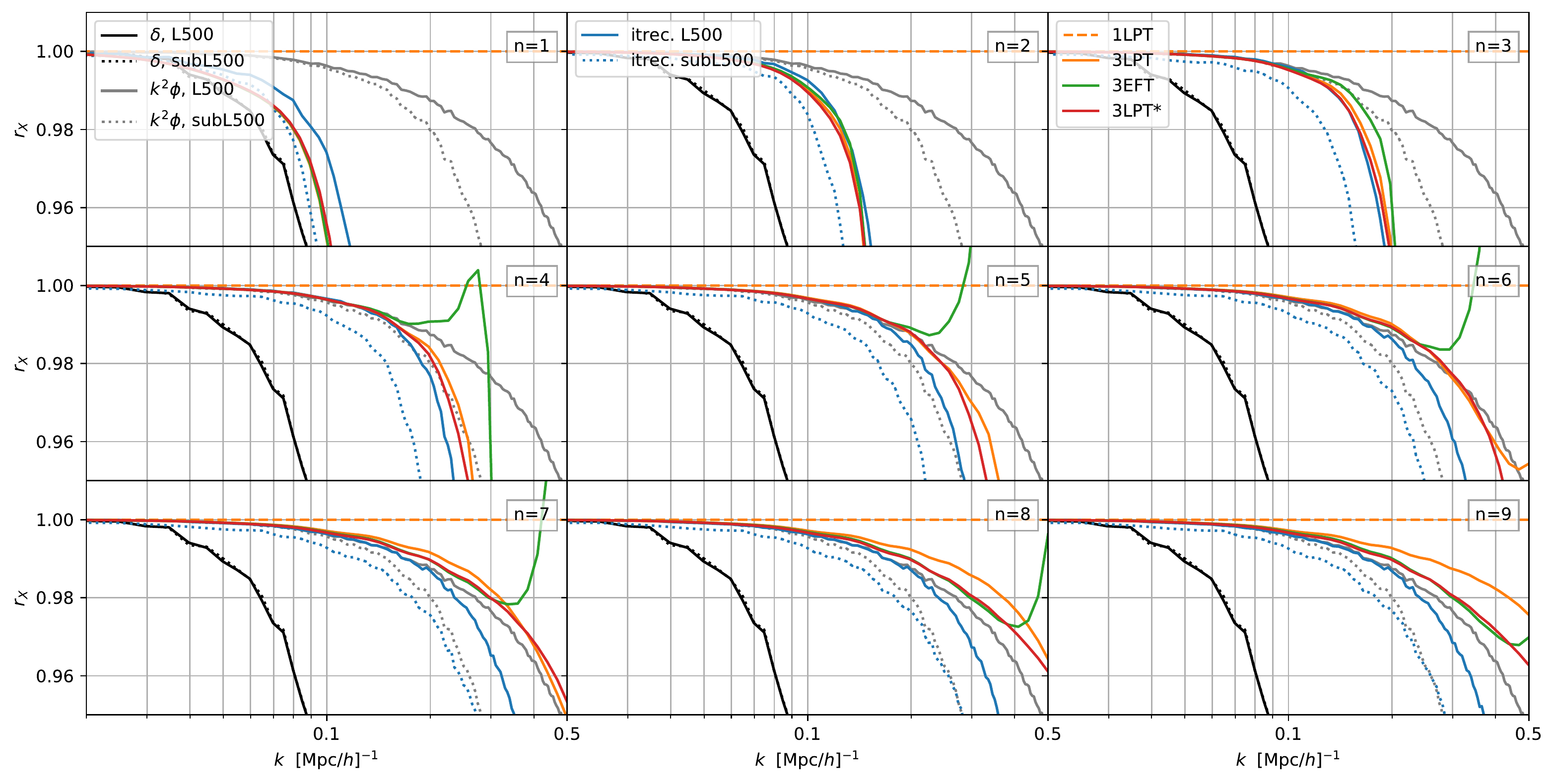}
  \caption{The cross-correlation coefficient counterparts of Fig.~\ref{1-loopres1}.
  3LPT, 3EFT, the true displacement and the simulation post reconstruction displacement agree well, while $r=1$ for 1LPT, i.e., the linear perturbations.
  While the cross-correlation coefficient should be always bounded by 1, the EFT is divergent for high $k$ since we introduced $-k^2$ term.  
  }
  \label{1-loopres3.1}

  \includegraphics[width=\linewidth]{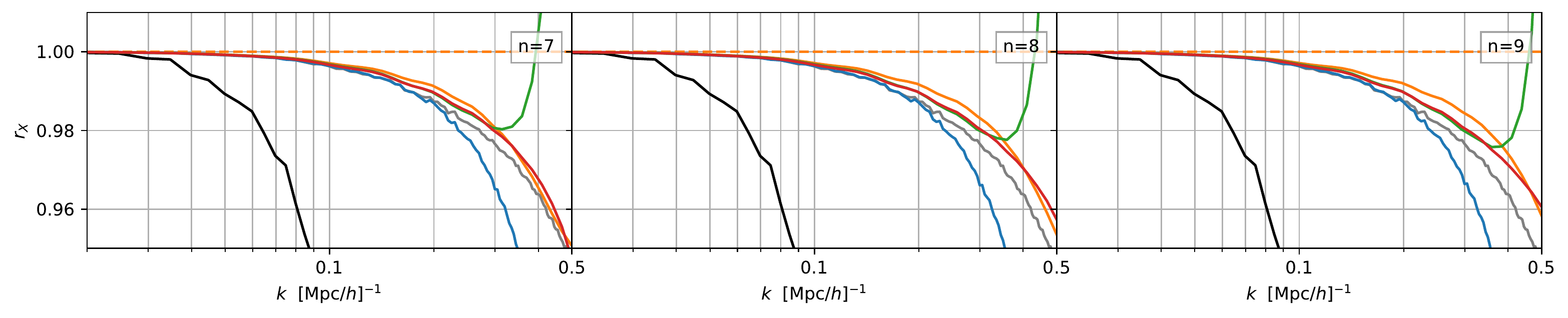}
  \caption{The cross correlation coefficient in Fig.~\ref{1-loopres1.1} with the minimum smoothing scale $R_{\rm min}\sim 3.5$Mpc/$h$. Plots are the same with Fig.~\ref{1-loopres3.1} up to $n=6$.}
  \label{1-loopres3.2}
\end{figure*}

In this section, we compare our LPT model derived in Sec.~\ref{sec5} with the simulation result and check if the discrepancy between the iterative reconstruction and the displacement field can be explained. 
Using Eqs.~\eqref{1-loop:power} and \eqref{1-loop:prop} we obtained the solid orange curves in Figs.~\ref{1-loopres1},~\ref{1-loopres1.1}, and \ref{1-loopres3.1}.
For $n< 4$, except for the first step, the 1LPT (dashed orange curves) and 3LPT (solid orange curves) solutions are very similar, and they reproduce the simulation result (solid blue lines) in Figs.~\ref{1-loopres1},~\ref{1-loopres1.1}, and~\ref{1-loopres3.1}. That is, the model explains the discrepancy between the postreconstruction field and the displacement field (gray lines) properly.
For higher iterations, LPT predictions begin to deviate, and only the 3LPT can predict the excess above the linear power spectrum, which is reasonable as we are trying to reconstruct smaller scale information that requires a higher-order description.
Nevertheless, we overall observe that the LPT models do not describe the numerical results very well. Looking at the propagator (Fig.~\ref{1-loopres1.1})  and the cross-correlation coefficient (Fig.~\ref{1-loopres3.1}), we find that the main deviation happens in the propagator, while the prediction of the cross-correlation coefficient is in excellent agreement for $n < 6$ and within less than 1\% at $k{\rm Mpc}/h \sim 0.2$ even at $n=9$. 

On the other hand, we find that the 3LPT up to $n = 6$ is converging to the true displacement obtained in the simulation. The 3LPT model for $n>6$ then deviates from the true displacement field to the 3LPT model of the displacement field. This shows that the perturbation theory predicts that the iterative process we adopted should recover the true displacement field up to $k{\rm Mpc}/h\lesssim 0.2$.
 The relatively weak convergence of the simulated post-iterative reconstruction field to the true nonlinear displacement field after $n=4$ implies the iterative reconstruction simulation includes some unknown numerical or technical effect that is not described in Sec.~\ref{stdbao} and \ref{itrbao}, particularly in the propagator.
Given that the offset between the model and the simulations is similarly parametrized as Eq.~\eqref{tobias_eft}, we conduct a quick test of an EFT model by simply adding a counter term
\begin{align}
	P^{{\rm 3EFT}}_{\phi^{(n)}_{\rm rec}} = P^{{\rm 3LPT}}_{\phi^{(n)}_{\rm rec}} + 2\alpha k^2 P^{{\rm 1LPT}}_{\phi^{(n)}_{\rm rec}},
	\label{3EFT}
\end{align}
and show the green curves in Figs.~\ref{1-loopres1},~\ref{1-loopres1.1}, and \ref{1-loopres3.1}. $\alpha$ is the same as for the displacement field. This test indicates that the offset between the model and the reconstructed field in the propagator is less likely mitigated by a simple counter term, and again the perturbation theory model expects that the iterative reconstruction converges to the displacement field in the end.

Another likely source of the discrepancy is the aforementioned numerical noise/artifact. Earlier, we justified subsampling particles (i.e., L500) by inspecting the convergence of the displacement field measurement on the L1500  (Sec.~\ref{simsec}).
However, the convergence in the measurements does not necessarily mean convergence in the process of reconstruction. 
The UV sensitivity of the loop integral implies that the displacement field in a simulation is sensitive to the particle resolution/density, and the effect may have been more complicated/more significant in the backward evolution of reconstructing the field based on the subsampled tracers in the iterative reconstruction. Our numerical resource is limited conducting the iterative reconstruction on fullL1500, but we present the case of 0.15\% subsampling (subL500) in Figs.~\ref{1-loopres1},~\ref{1-loopres1.1}, and \ref{1-loopres3.1}, as dotted lines to show the greater sensitivity to the particle subsampling in the process of reconstruction compared to the displacement field. Including a shot noise effect in the theoretical model and implementing a pixel window function de-convolution in the simulated field will be crucial for further comparison, which we plan to investigate in a future paper.

To account for the expected breakdown due to various missing small-scale physics of our subsampled catalog, we manually impose a minimum smoothing scale in the reconstruction in an attempt to control such an effect before reaching this breakdown. 
Fig.~\ref{fig222} showed that the reconstruction changes the convergence behavior around $n=6$, which is suggestive of the effective minimum smoothing scale in the simulation and the theory to be the smoothing scale around this step, which is $3.5.{\rm Mpc}/h$.
Therefore, instead of Eq.~\eqref{Rdef}, we set
\begin{align}
R_n =  &{\rm max}(\epsilon^{n} R,R_{\rm min}),
\end{align}
with $R_{\rm min}=3.5.{\rm Mpc}/h$ and show the results in Figs.~\ref{1-loopres2}, \ref{1-loopres2.1} and \ref{1-loopres3.2}.
Introducing the minimum smoothing scale derives a convergence at $n=7$, as we cannot extract the information from smaller scales than the smoothing scale. However, the discrepancy between the theory and the simulation remains.

Based on the lack of UV modes that we cannot explain, we introduce a phenomenological ad hoc model that we call 3LPT*.
In this model, we subtract all UV sensitive integrals from the 3LPT power spectrum and define 
\begin{align}
	&P^{{\rm 3LPT*}}_{\phi^{(n)}_{\rm rec}}\equiv 	 
	\bar A^{(n)2}\left [ 
	P^{\rm 1LPT}_{\phi}
	+
	P_{\phi22}
	\right]
	+
	P^{{\rm 3LPT}}_{\phi^{(n)}_{\rm rec}22},\label{1-loop:power22}
\end{align}	
and we show the red curves in Figs.~\ref{1-loopres1},~\ref{1-loopres1.1}, and \ref{1-loopres3.1} and Figs.~\ref{1-loopres2}, \ref{1-loopres2.1} and \ref{1-loopres3.2}. This model takes the propagator of 1LPT, which better describes the observed propagator of the reconstructed field than 3LPT.  
As a caveat, there is no reasonable explanation about this ad hoc truncation, and this is an inconsistent treatment as either loop expansion or perturbative expansion.
However, we found the 3LPT* fits the simulated iterative reconstruction at 1\% accuracy for $n>6$ for both cases with and without the minimum smoothing scale.

\section{Conclusions}\label{secdisc}
As a recent study~\cite{Baldauf:2015tla} has pointed out, the nonlinear displacement at low redshift maintains a high cross-correlation with the initial field, contrary to the corresponding nonlinear density field. 
Thus, if one could measure the nonlinear displacement field directly in the late time Universe, the measurement would contain the BAO feature almost without the nonlinear degradation, and it could also allow us to consistently use the information beyond the BAO, i.e., the full clustering shape, based on nonlinear perturbation theory.  

Ref. \cite{Schmittfull:2017uhh} then developed a  non-standard extension to the density reconstruction technique, which was designed to reproduce the displacement field from the observed nonlinear density fluctuations by iteratively moving mass tracers along the gradient of the progressively smoothed density field to the Lagrangian positions. 
This paper investigated how closely the method returned the nonlinear displacement field, focusing on the BAO and the full shape. Then, we derived an LPT-based model for the reconstructed displacement field and tested the model against the simulated reconstructed displacement field.  We summarize our main results below.

First, we confirmed that the displacement field is highly correlated with the linear field; the degradation effect and the phase shift on the BAO are negligible, and the nonlinear instability is milder than the nonlinear density field. However, the full shape of the nonlinear displacement field is different from the linear matter power spectrum on the BAO scale by $\sim 8\%$, and modeling with a higher-order perturbation theory with a correction term for the UV sensitive loops, such as the EFT model, is required for precision cosmology.

We found that the postreconstructed displacement field approaches closer to the nonlinear displacement field with increasing iterations but does not perfectly recover the nonlinear displacement field in the end. There is an 8\% discrepancy between the true displacement field and the simulated postreconstruction displacement field at $k{\rm Mpc}/h\sim 0.2$ due to unknown effects specific to the iterative process in the simulation. The discrepancy is mainly in the propagator (i.e., the shift term), while the cross-correlation coefficient (i.e., mode-coupling term) agrees well.
At the final iteration, the postreconstruction field tends to converge closer to the linear power spectrum. 

We built perturbation theory models for the reconstructed field to understand the difference.
In this process, we realized that the UV mistake should largely cancel for the reconstructed displacement for moderate iterations in the process of estimating the displacement field, allowing us to avoid introducing the EFT counter term up to $k{\rm Mpc}/h\lesssim 0.2$ at $z=0.6$.  We, therefore, adopt the 1-loop LPT approach.

We modeled the iterative reconstruction as follows: firstly, we put forward the $n$th reconstruction step displacement field ansatz in the most general form, expanding the equations up to third order in the 0th step, i.e., the nonlinear displacement field corresponding to the observed nonlinear density. Then we derived the $(n+1)$th step displacement field and obtained the recurrence relation for the expansion kernels.
This allows us to systematically compute an arbitrary step's postreconstruction 1-loop spectrum using the recurrence relation.

Our model predicts that the postreconstruction displacement field should converge to the true displacement field, contrary to what we found in our simulation. The model provides a good description of the cross-correlation coefficient, while the difference in the propagator is mainly responsible for the discrepancy between the model and the simulation.

While we could not identify the reason for the discrepancy between the model and the reconstructed field in the present paper, we suspect this is potentially related to the subsampling of the tracers. Including such an effect in the theoretical model and implementing a pixel window function correction will help identify the discrepancies. We leave such an extension for future papers. 

Our tentative solution to the discrepancy is to drop all UV-sensitive integrals from the 3LPT spectrum, and the ad hoc model fits the postreconstruction displacement field with 1\% accuracy for $n>6$.

This paper presents the proof of concept on how we can model the iterative reconstruction steps using the real-space mass distribution with little shot noise effect.
Improving the agreement between the model and the simulation in future papers, we expect it would be straightforward to generalize our modeling to the redshift space. 
On the other hand, an extension to biased tracers would be nontrivial, based on our investigation in the companion paper~\cite{Seo:2021}.  While the model we built appears quite strenuous, we note that all iteration kernels are cosmology-independent and could be precomputed and interpolated. A short come is that we find that the speed of 1-loop calculation is still slow. We confirmed that the prereconstruction 3LPT calculation could be optimized with FFTlog, using the technique presented in Ref.~\cite{Schmittfull:2016jsw}. However, we found that the same algorithm does not apply to the postreconstruction spectrum because of the complex structure of the iteration kernels.
Thus, the iterative reconstruction requires 2D standard quadrature integration. Speeding up those calculations would be critical for actual data analysis. Additionally, it would be essential to extract a more compressed and practical form of the model that can be more efficiently utilized for data analysis, which we plan to do to implement the redshift space distortions and galaxy bias.

Finally, we note that our work is still in an early stage compared to the modeling of the standard reconstruction density field using the 1-loop SPT in Ref.~\cite{Hikage:2017tmm} and the Zeldovich approximation in Ref.~\cite{Chen:2019lpf}, etc. Moreover, Ref.~\cite{Seo:2021} implies that the iterative displacement reconstruction needs to be better optimized in the presence of high shot noise to take full advantage of it. We nevertheless believe that reconstructing the nonlinear displacement field could be pretty helpful in a deficient shot noise regime that can be available in future galaxy surveys due to its high correlation with the initial condition and, therefore, worth further investigation. 

\begin{acknowledgments}

The authors would like to thank Marcel Schmittfull for providing simulations and valuable discussions.
The authors also would like to thank Stephen Chen for their helpful discussions.
AO would like to thank Masaru Hongo for valuable discussions. AO and H.-J.S. are supported by the U.S.~Department of Energy, Office of Science, Office of High Energy Physics under DE-SC0019091. 
SS was supported in part by World Premier International Research Center Initiative (WPI Initiative), MEXT, Japan. 
This project has received funding from the European Research Council (ERC) under the European Union's Horizon 2020 research and innovation program (grant agreement 853291). FB is a University Research Fellow.

\end{acknowledgments}

\appendix
\section{Recurrence relation in the explicit form}\label{woind}
Eqs.~\eqref{rec1_}, \eqref{rec2_2} and \eqref{rec3_3} are reduced to the following expressions without index:
\begin{widetext}
\begin{align}
A^{(n+1)}(\mathbf k_1) &= A^{(n)}(\mathbf k_1)- S(k_1)A^{(n)}(\mathbf k_1) \label{Eq.A1}
\\
A^{(n+1)}(\mathbf k_1,\mathbf k_2) &=A^{(n)}(\mathbf k_1,\mathbf k_2) 
	-S(k)A^{(n)}(\mathbf k_1,\mathbf k_2)  -S^{(n)}(k)
	\frac{(\mathbf k \cdot \mathbf k_1)(\mathbf k \cdot \mathbf k_2)}{k^2} 
	A^{(n)}(k_1) A^{(n)}(k_2) 
	\notag \\
	&+\frac{ ( \mathbf k_1\cdot \mathbf k_2)}{k^2}\left[  (\mathbf k\cdot \mathbf k_2) S^{(n)}(k_2)+ (\mathbf k\cdot \mathbf k_1) S^{(n)}(k_1) \right]
    A^{(n)}(k_1)  A^{(n)}(k_2),\label{A2}
    \\
	A^{(n+1)}(\mathbf k_1,\mathbf k_2,\mathbf k_3) &=A^{(n)}(\mathbf k_1,\mathbf k_2,\mathbf k_3)
    -S^{(n)}(k)A^{(n)}(\mathbf k_1,\mathbf k_2,\mathbf k_3)
     \notag \\
    &
      -  \frac{ (\mathbf k \cdot \mathbf k_1)(\mathbf k \cdot \mathbf k_2)(\mathbf k \cdot \mathbf k_3)}{k^2} S^{(n)}(k) A^{(n)}(\mathbf k_1)A^{(n)}(\mathbf k_2)A^{(n)}(\mathbf k_3)
	\notag \\
	&
	+\Bigg[-S^{(n)}(k) \frac{(\mathbf k \cdot \mathbf k_{12})(\mathbf k \cdot \mathbf k_3)}{k^2}  A^{(n)}(\mathbf k_1,\mathbf k_2) A^{(n)}(\mathbf k_3)  
    \notag \\
    &
   +   \frac{ ( \mathbf k_{12}\cdot \mathbf k_{3})(\mathbf k\cdot \mathbf k_3)}{k^2} S^{(n)}(k_3)   A^{(n)}(\mathbf k_1,\mathbf k_2) A^{(n)}(\mathbf k_3) 
   \notag \\
	&   +
     \frac{ ( \mathbf k_{3}\cdot \mathbf k_{12})(\mathbf k\cdot \mathbf k_{12})}{k_a^2} S^{(n)}(k_{12}) A^{(n)}(\mathbf k_1,\mathbf k_2) A^{(n)}(\mathbf k_3)  
   	\notag \\
	&+
   \frac{(\mathbf k\cdot \mathbf k_{23})( \mathbf k_{23}\cdot \mathbf k_{1})(\mathbf k_{23} \cdot \mathbf k_2)(\mathbf k_{23} \cdot \mathbf k_3)}{k^2k_{23}^2} S^{(n)}(k_{23})
	A^{(n)}(\mathbf k_1)A^{(n)}(\mathbf k_2)A^{(n)}(\mathbf k_3) 
   \notag \\
      &  
- \frac{(\mathbf k_{1}\cdot \mathbf k_3)(  \mathbf k_{2} \cdot  \mathbf   k_3)(  \mathbf k \cdot  \mathbf   k_3)}{k^2} S^{(n)}(k_3) A^{(n)}(\mathbf k_1)A^{(n)}(\mathbf k_2)A^{(n)}(\mathbf k_3) +{\rm 2~perms.}\Bigg],\label{rec3_3_s}
\end{align}
where $\mathbf k = \mathbf k_1+\mathbf k_2+\mathbf k_3$, and $\mathbf k_{ij}=\mathbf k_i+\mathbf k_j$.
\end{widetext}

\bibliography{bib}{}
\bibliographystyle{unsrturl}

\end{document}